\documentclass[english,reprint,a4paper,aps,prb]{revtex4-1}
\usepackage[T1]{fontenc}
\usepackage[latin9]{inputenc}
\usepackage{amsmath}
\usepackage{graphicx}
\usepackage{esint}

\makeatletter

\@ifundefined{textcolor}{}
{%
 \definecolor{BLACK}{gray}{0}
 \definecolor{WHITE}{gray}{1}
 \definecolor{RED}{rgb}{1,0,0}
 \definecolor{GREEN}{rgb}{0,1,0}
 \definecolor{BLUE}{rgb}{0,0,1}
 \definecolor{CYAN}{cmyk}{1,0,0,0}
 \definecolor{MAGENTA}{cmyk}{0,1,0,0}
 \definecolor{YELLOW}{cmyk}{0,0,1,0}
}

\makeatother

\usepackage{babel}
\begin{document}

\title{Coagulation kinetics beyond mean field theory using an optimised
Poisson representation}

\author{James Burnett$^{*}$ and Ian J. Ford$^{\dagger}$ }

\affiliation{$^{*}$Department of Mathematics, UCL, Gower Street, London WC1E
6BT, United Kingdom }

\affiliation{$^{\dagger}$Department of Physics and Astronomy, UCL, Gower Street,
London WC1E 6BT, United Kingdom}
\begin{abstract}
Binary particle coagulation can be modelled as the repeated random
process of the combination of two particles to form a third. The kinetics
may be represented by population rate equations based on a mean field
assumption, according to which the rate of aggregation is taken to
be proportional to the product of the \emph{mean} populations of the
two participants, but this can be a poor approximation when the mean
populations are small. However, using the Poisson representation it
is possible to derive a set of rate equations that go beyond mean
field theory, describing pseudo-populations that are continuous, noisy
and complex, but where averaging over the noise and initial conditions
gives the mean of the physical population. Such an approach is explored
for the simple case of a size-independent rate of coagulation between
particles. Analytical results are compared with numerical computations
and with results derived by other means. In the numerical work we
encounter instabilities that can be eliminated using a suitable `gauge'
transformation of the problem {[}P. D. Drummond, Eur. Phys. J. B38,
617 (2004){]} which we show to be equivalent to the application of
the Cameron-Martin-Girsanov formula describing a shift in a probability
measure. The cost of such a procedure is to introduce additional statistical
noise into the numerical results, but we identify an optimised gauge
transformation where this difficulty is minimal for the main properties
of interest. For more complicated systems, such an approach is likely
to be computationally cheaper than Monte Carlo simulation.
\end{abstract}
\maketitle

\section{Introduction}

The process of coagulation or aggregation is responsible for the coarsening
of a size distribution of particles suspended in gaseous or liquid
media. The phenomenon consists of a sequence of statistically independent
events where two (or possibly more) particles unite, perhaps as a
result of collision, to create a composite particle, and each event
reduces the number of particles in the system \cite{Smoluchowski06,Leyvraz03,Lushnikov06}.
This has practical consequences such as colloidal precipitation or
accelerated rainfall from clouds \cite{Pruppacher97,Mehlig04,McGraw06}.
Fragmentation can take place as well \cite{Spicer96}, but precipitation
processes are typically dominated by irreversible agglomeration. The
phenomenon is familiar and yet it can present some surprises, an example
of which was presented by Lushnikov \cite{Lushnikov78} in an exact
study of coagulation kinetics driven by various choices of aggregation
rates. He showed that in the later stages of a process where the binary
coagulation rate is proportional to the product of the masses of the
participants, a single particle emerges with a mass representing a
sizable fraction of that of the entire system \cite{Lushnikov05}.
This has been termed a gelation event and standard kinetic models
of coagulation are unable to account for the phenomenon, for the simple
reason that they are designed to describe systems consisting of very
large populations of particles in each size class. They rely on a
mean field approximation, though this is not always clearly recognised.
When small particle populations play a key role in the kinetics, different
approaches become necessary \cite{Krug03}, the most common being
Monte Carlo modelling \cite{Smith98}, though analytic treatments
are sometimes possible \cite{Hendriks85}.

This paper investigates the utility of a rate equation model of kinetic
processes that is able to capture small population effects. The Markovian
master equations that describe coagulation may be transformed mathematically
into a problem in the dynamics of continuous stochastic variables
acted upon by complex noise, the solution to which may be related
to the evolving statistical properties of the populations in the physical
system. The recasting of the problem into one that concerns complex
`pseudo-populations' can be done in two distinct ways, either using
methods of operator algebra similar to those employed in quantum field
theories \cite{Doi76a,Doi76b,Peliti85,Patzlaff94,Rey97,Mattis98,Sasai03,Tauber05,Ohkubo07,Schulz08,Ohkubo11},
or through the so-called Poisson representation of the populations
\cite{Gardiner77,Gardiner09,Drummond04,Drummond10}. The physical
problem concerning the stochastic evolution of a set of (integer-valued)
populations is replaced by the task of solving and averaging a set
of stochastic differential equations \cite{Deloubriere02,Hochberg06}.
The purpose of this paper is to explore the use of the Poisson representation
in this area.

The advantage of such a transformation over Monte Carlo simulation
of the process \cite{Gillespie77,Smith98} emerges for cases where
the particles come in many species or sizes, since the configuration
space of the populations, and hence the computational cost, increases
very rapidly, but in order to illustrate the approach we study a very
simple example of coagulation, where the aggregation rates do not
depend on the masses of the participants. This is very different from
the cases that exhibit gelation, studied by Lushnikov \cite{Lushnikov78}
and others \cite{Hendriks85}, and we might not expect major deviations
from a mean field approach, but nevertheless it is possible to use
the system to demonstrate the analytic procedure, and compare the
accuracy of numerical pseudo-population dynamics and averaging with
respect to other approaches \cite{Barzykin05,Biham01,Lushnikov03,Bhatt03,Green01,LosertV-K07},
in order to form a view on the usefulness of the approach. We do not
suggest that such a method is to be preferred over others for such
simple examples, since the technique is much more elaborate. There
might, nevertheless, be cases where the approach might have analytic
advantages even where mean field equations are largely adequate, and
the area of nucleation modelling and barrier crossing is one worth
exploring \cite{Bhatt03,Yvinec12}.

In Section \ref{sec:Analytic-model-of} the master equations describing
the basic problem of $A+A\to A$ aggregation are developed and the
Poisson representation is introduced and used to derive the equivalent
stochastic dynamics problem and associated averaging scheme. The evolving
complex pseudo-population is obtained and its properties established.
In Section \ref{sec:Numerical-approach.} a parallel numerical study
of the stochastic problem is discussed. Inherent instabilities in
the dynamics may be eliminated using the Cameron-Martin-Girsanov formula,
or equivalently through a gauge transformation introduced by Drummond
\cite{Drummond04}. However, this comes with the introduction of diffusive
dynamics for a subsidiary variable that imposes a cost on computational
accuracy. Nevertheless, with a certain choice of transformation, termed
an optimised gauge, we can ensure that the coagulation process is
completed before this diffusive noise becomes apparent. We comment
on the procedures and prospects for their further use in more complicated
models in Section \ref{sec:Discussion}.

\section{Analytic model of coagulation\label{sec:Analytic-model-of}}

\subsection{Master equations}

We consider the dynamics of a population of particles of a species
$A$ undergoing binary coagulation $A+A\rightarrow A$. The distinction
between particles of different mass is ignored and the rate at which
aggregation events takes place is a constant, making this one of the
simplest cases to study. The evolution of $P(N,t)$, the probability
at time $t$ that $N$ particles remain, is described by a set of
master equations of the form
\begin{equation}
\frac{dP(N,t)}{dt}=\kappa(N+1)NP(N+1,t)-\kappa N(N-1)P(N,t).\label{eq:100}
\end{equation}
The first term corresponds to the gain in the probability of observing
population $N$ as a result of an aggregation event amongst $N+1$
particles, weighted by the number of particle pairs in such a state
and the probability $\kappa/2$ of an event per unit time and per
pair. The second term corresponds to the loss of probability due to
aggregation starting from a population equal to $N$. Multiplying
the master equations by $N$ and summing gives
\begin{equation}
\frac{d\overline{N}}{dt}=-\kappa\left(\overline{N^{2}}-\overline{N}\right),\label{eq:101}
\end{equation}
where statistical averages are written $\overline{F(N)}=\sum_{0}^{\infty}F(N)P(N,t)$.

If we take the view that $\overline{N^{2}}\approx\overline{N}^{2}$,
which corresponds to the neglect of fluctuations in population (a
mean field approximation), and consider $\overline{N}\gg1$, such
that only the first term on the right hand side in Eq. (\ref{eq:101})
is retained, we can write
\begin{equation}
\frac{d\overline{N}}{dt}\approx-\kappa\overline{N}^{2},\label{eq:101a}
\end{equation}
leading to
\begin{equation}
\overline{N}\approx\frac{\overline{N}_{0}}{1+\kappa\overline{N}_{0}t},\label{eq:102}
\end{equation}
where $\overline{N}_{0}$ is the initial mean population. This is
the well known Smoluchowski solution to this type of coagulation kinetics
\cite{Smoluchowski06}.

If the second term in Eq. (\ref{eq:101}) is retained, but again neglecting
fluctuations, then the integration yields
\begin{eqnarray}
\overline{N} & \approx & \frac{\overline{N}_{0}\exp(\kappa t)}{1-\overline{N}_{0}\left(1-\exp(\kappa t)\right)},\label{eq:103}
\end{eqnarray}
which ought to be more accurate than Eq. (\ref{eq:102}), especially
in the limit $t\to\infty$ when it goes to unity rather than zero.

Rather than making the mean field approximation $\overline{N^{2}}\approx\overline{N}^{2}$,
we could generate an evolution equation for $\overline{N^{2}}$ by
multiplying the master equation by $N^{2}$ and summing, namely $d\overline{N^{2}}/dt=\kappa(-2\overline{N^{3}}+3\overline{N^{2}}-\overline{N})$,
but now the third moment appears on the right hand side. An equation
for the evolution of this moment would involve the fourth moment:
the hierarchy problem that often arises in kinetic theory. A closure
condition such as $\overline{N^{3}}=\overline{N^{2}}\,\overline{N}$
could be imposed, but the accuracy would be questionable.

Similarly, the numerous master equations for the $P(N_{1},N_{2},\dots,t)$
in a more general model where there are different categories of species
$\left\{ A_{i}\right\} $ with populations $\left\{ N_{i}\right\} $
would reduce under a mean field approximation to
\begin{equation}
\frac{d\overline{N_{i}}}{dt}=\frac{1}{2}\sum_{j=1}^{i-1}\kappa_{i-j,j}\overline{N}_{i-j}\overline{N}_{j}-\sum_{j}\kappa_{i,j}\overline{N}_{i}\overline{N}_{j},\label{eq:105}
\end{equation}
where the coefficients $\kappa_{i,j}$ quantify the rate of aggregation
between species $i$ and $j$, with $i$ representing species mass,
for example. If we should need a treatment beyond the mean field approximation,
then for relatively simple stochastic processes we could use Monte
Carlo simulations to extract the relevant statistical properties,
but for more complex coagulation problems the use of this technique
can become computationally expensive.

We investigate a transformation of the problem that reduces the master
equations to the form of Eq. (\ref{eq:101}) and more generally Eq.
(\ref{eq:105}), but supplemented by a noise term on the right hand
side. This is not the same as inserting noise to represent a random
source term in the population dynamics, nor is it an additional representation
of stochasticity in the coagulation kinetics. It is a noise term that
creates the statistical correlations in the populations that are neglected
when the mean field approximation is taken. It turns out that the
dynamics of simple binary coagulation may be modelled by a pseudo-population
$\phi$ that evolves according to the equation
\begin{equation}
\frac{d\phi}{dt}=-\kappa\phi^{2}+\xi(t),\label{eq:106}
\end{equation}
where $\xi$ is a noise with certain statistical properties. This
may be compared with Eq. (\ref{eq:101a}). The notable aspect of this
representation is that the noise is complex, such that the variable
$\phi$ is generally complex as well. Averages are to be taken over
the noise history to make a connection with the dynamics of a real
population. Nevertheless, solving Eq. (\ref{eq:106}) can be a simpler
task than that posed by Eq. (\ref{eq:100}). We next describe how
this transformation may be achieved.

\subsection{Poisson representations\label{sub:Poisson-representations}}

Gardiner and Chaturvedi \cite{Gardiner77}\emph{ }outlined a method
for transforming master equations into a Fokker-Planck equation by
representing probability distributions as integrals of weighted complex
Poisson distributions. The probability of finding $N$ particles at
time $t$ may be written
\begin{equation}
P(N,t)=\oint_{C}f(\phi,t)\mathrm{exp}(-\phi)\frac{\phi^{N}}{N!}d\phi.\label{eq:107}
\end{equation}
This is a superposition of Poisson distributions, over a closed contour
of complex mean values $\phi$, with an evolving weighting function
$f(\phi,t)$ which has initial value
\begin{equation}
f(\phi,0)=\sum_{N}P(N,0)\frac{N!}{2\pi i}\phi^{-N-1}\exp(\phi),\label{eq:108}
\end{equation}
such that if the contour $C$ includes the origin, the initial condition
$P(N,0)$ is recovered from Eq. (\ref{eq:107}). A Poisson representation
is clearly one of a number of approaches where a probability distribution
is written as an integral transform of some weighting function with
respect to a chosen kernel. Representations used in some familiar
generating function methods come into this category, and a very close
relation is the inverse $z$-transform that involves an analogue of
Eq. (\ref{eq:107}) with a power-law kernel rather than a Poisson
distribution. However, the latter is rather suitable for the problem
under consideration here, partly because such distributions often
emerge as solutions to growth problems, but also because the form
of the derived equation for the evolution of the weighting function
has some intuitive value.

There are two particular initial conditions of interest. If the initial
population is known to be $N_{0}$ then $P(N,0)=\delta_{NN_{0}}$
and
\begin{equation}
f(\phi,0)=f_{N_{0}}(\phi)=\frac{N_{0}!}{2\pi i}\phi^{-N_{0}-1}\mathrm{exp}(\phi),\label{eq:109}
\end{equation}
whereas if $P(N,0)$ is a Poisson distribution with mean $\lambda$
then we would use
\begin{equation}
f(\phi,0)=f_{\lambda}^{P}(\phi)=\sum_{N}\frac{1}{2\pi i}\frac{\lambda^{N}}{\phi^{N+1}}\exp\left(\phi-\lambda\right).\label{eq:110}
\end{equation}
Alternatively, we could employ a Poisson representation that involves
a 2-d integration over the entire complex plane \cite{Gardiner77},
namely
\begin{equation}
P(N,t)=\int f(\phi,t)\mathrm{exp}(-\phi)\frac{\phi^{N}}{N!}d^{2}\phi,\label{eq:110a}
\end{equation}
which is particularly convenient if $P(N,t)$ is initially a Poisson
distribution since we would then use $f(\phi,0)=\delta(\phi-\lambda)$.

Substituting the Poisson representation of the probabilities given
in Eqs. (\ref{eq:107}) or (\ref{eq:110a}) into Eq. (\ref{eq:100})
leads to an evolution equation for $f(\phi,t)$:
\begin{equation}
\frac{\partial f}{\partial t}=\kappa\frac{\partial(\phi^{2}f)}{\partial\phi}-\kappa\frac{\partial^{2}(\phi^{2}f)}{\partial\phi^{2}}.\label{eq:111}
\end{equation}
This emerges as long as integration boundary terms can be dropped,
which could potentially be a problem for the 2-d integration scheme
Eq. (\ref{eq:110a}) but is not an issue for the closed contour integral
representation Eq. (\ref{eq:107}) \cite{Gardiner09}. And since a
stochastic problem described by a Fokker-Planck equation (FPE) such
as (\ref{eq:111}) can be recast in terms of a stochastic differential
equation (SDE), the properties of the distribution $f(\phi,t)$ can
be reconstructed by solving
\begin{equation}
d\phi=-\kappa\phi^{2}dt+i(2\kappa)^{1/2}\phi dW_{t},\label{eq:112}
\end{equation}
where $dW_{t}$ is an increment in a Wiener process with properties
$\langle dW_{t}\rangle=0$ and $\langle(dW_{t})^{2}\rangle=dt$, the
brackets representing an average over the noise. This Ito-type SDE
takes the promised form of Eq. (\ref{eq:106}). Note that the noise
term is complex because the second derivative in Eq. (\ref{eq:111})
has a negative coefficient. The equivalence between FPE and SDE approaches
is such that the average of a function $\hat{A}(\phi)$ weighted according
to the solution $f(\phi,t)$ with initial condition $f(\phi,0)=\delta(\phi-\phi_{0})$
(namely the Green's function of the FPE) is equal to the average of
the same function of a stochastic variable $\phi(t)$ evolved over
a noise history $W_{t}$ with initial condition $\phi(0)=\phi_{0}$.
Both averages will be denoted with angled brackets and subscript in
the form $\langle\hat{A}(\phi)\rangle_{\phi_{0}}$ with time dependence
understood.

It is possible to solve Eq. (\ref{eq:112}) exactly. The details are
given in Appendix \ref{sec:Stochastic-evolution-of}, and the solution
is
\begin{equation}
\phi(t)=\frac{\phi_{0}\mathrm{exp}(\kappa t+i(2\kappa)^{\frac{1}{2}}W{}_{t})}{1+\kappa\phi_{0}\int_{0}^{t}\mathrm{exp}(\kappa s+i(2\kappa)^{\frac{1}{2}}W_{s})ds}.\label{eq:119m}
\end{equation}
We next show that an average of this stochastic variable can be related
to $\overline{N}$, the particle population averaged over the stochasticity
of the\emph{ }physical\emph{ }coagulation process.

\subsection{Integrals over initial pseudo-population}

Using Eq. (\ref{eq:107}) we can write
\begin{eqnarray}
\overline{A(N)} & = & \sum_{N=0}^{\infty}A(N)\oint_{C}f(\phi,t)\exp(-\phi)\frac{\phi^{N}}{N!}d\phi\nonumber \\
 & = & \oint_{C}f(\phi,t)\hat{A}(\phi)d\phi=\langle\hat{A}(\phi)\rangle,\label{eq:120}
\end{eqnarray}
noting that the final expression differs from the average $\langle\hat{A}(\phi)\rangle_{\phi_{0}}$
introduced in Section \ref{sub:Poisson-representations}, since the
initial weighting $f(\phi,0)$ takes the general form of Eq. (\ref{eq:108})
instead of $\delta(\phi-\phi_{0})$, and where
\begin{equation}
\hat{A}(\phi)=\sum_{N}A(N)\exp(-\phi)\frac{\phi^{N}}{N!}.\label{eq:121}
\end{equation}
For example, if $A(N)=N$, then $\hat{A}(\phi)=\phi$, and for $A(N)=N^{2}$
we have $\hat{A}(\phi)=\phi^{2}+\phi$. A similar construction might
be made using the 2-d integration scheme (\ref{eq:110a}). As for
the state probabilities $P(N,t)$, we note that $P(N^{\prime},t)=\sum_{N}\delta_{NN^{\prime}}P(N,t)=\overline{\delta_{NN^{\prime}}}$.
Taking $A(N)=\delta_{NN^{\prime}}$ we find that the corresponding
$\hat{A}(\phi)$ according to Eq. (\ref{eq:121}) is $\exp(-\phi)\phi^{N^{\prime}}/N^{\prime}!$
and hence
\begin{equation}
P(N,t)=\langle\exp(-\phi)\phi^{N}/N!\rangle.\label{eq:78}
\end{equation}

\begin{figure}
\begin{centering}
\includegraphics[width=1\columnwidth]{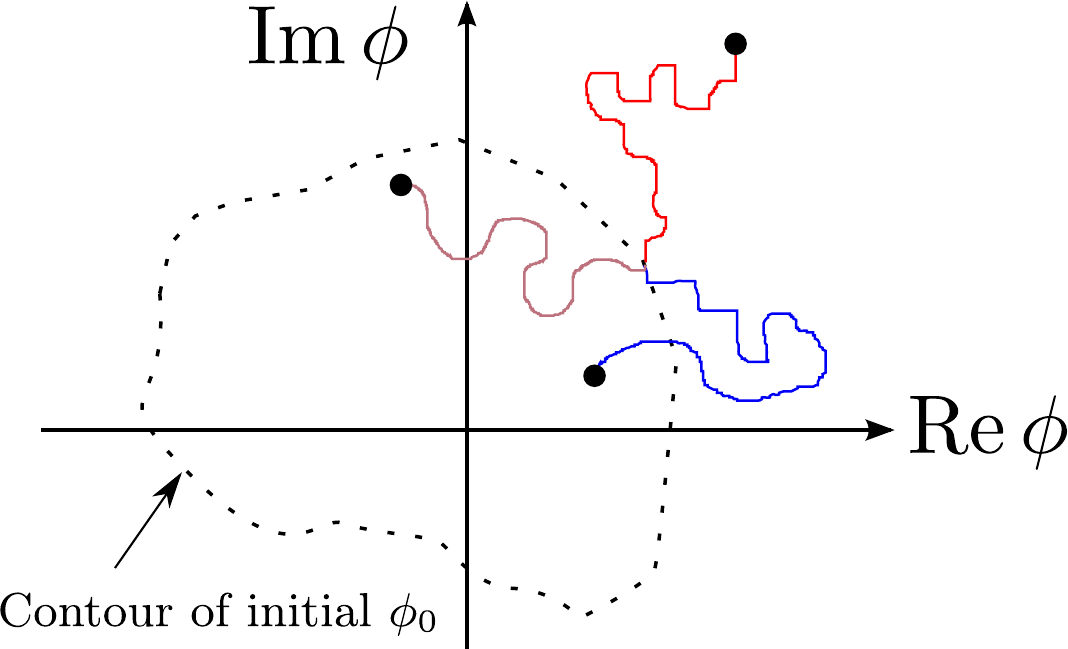}
\par\end{centering}

\protect\caption{Sketch of stochastic realisations of the evolution of pseudo-population
$\phi(t)$ in the complex plane according to Eq. (\ref{eq:112}),
starting from a point $\phi_{0}$ on a contour that encircles the
origin. An average over all end-points, three examples of which are
shown as filled circles, corresponds to the quantity $\langle\phi\rangle_{\phi_{0}}$.
\label{fig:Sketch-of-the}}
\end{figure}

As noted earlier, the Fokker-Planck equation (\ref{eq:111}), written
as $\mathcal{L}f(\phi,t)=0$, has a Green's function satisfying $\mathcal{L}G(\phi,\phi_{0},t)=0$
for $t>0$ with initial condition $G(\phi,\phi_{0},0)=\delta(\phi-\phi_{0})$
and we can write
\begin{equation}
\langle\hat{A}(\phi)\rangle_{\phi_{0}}=\oint_{C}G(\phi,\phi_{0},t)\hat{A}(\phi)d\phi,\label{eq:122}
\end{equation}
where the contour is obliged to pass through the point $\phi=\phi_{0}$
such that the initial condition $\langle\hat{A}(\phi)\rangle_{\phi_{0}}=\hat{A}(\phi_{0})$
is satisfied at $t=0$. Furthermore, the evolution of $f(\phi,t)$,
starting from an initial distribution $f(\phi,0)$ that is non-zero
only at points $\phi_{0}$ that define an origin-encircling contour
$C$, can be constructed from the superposition
\begin{equation}
f(\phi,t)=\oint_{C}d\phi_{0}G(\phi,\phi_{0},t)f(\phi_{0},0),\label{eq:123}
\end{equation}
so that from Eq. (\ref{eq:120}):
\begin{equation}
\overline{A(N)}=\oint_{C}\oint_{C}d\phi_{0}G(\phi,\phi_{0},t)f(\phi_{0},0)\hat{A}(\phi)d\phi,\label{eq:124}
\end{equation}
and we deduce using Eq. (\ref{eq:122}) that
\begin{equation}
\overline{A(N)}=\oint_{C}d\phi_{0}f(\phi_{0},0)\langle\hat{A}(\phi)\rangle_{\phi_{0}},\label{eq:125}
\end{equation}
which is intuitively understood as a superposition of the stochastic
evolution of a function of the pseudo-population $\phi(t)$ that evolves
from points on a contour of initial values $\phi_{0}$, weighted by
the function $f(\phi_{0},0)$, given in Eq. (\ref{eq:108}), and then
averaged over the noise present in its SDE (\ref{eq:112}) or in its
solution Eq. (\ref{eq:119m}). This picture is illustrated in Figure
\ref{fig:Sketch-of-the}.

We could equally well employ the representation $\overline{A(N)}=\int f(\phi,t)\hat{A}(\phi)d^{2}\phi$
together with the associated 2-d integral versions of Eqs. (\ref{eq:122})
and (\ref{eq:123}) to obtain an analogous result $\overline{A(N)}=\int d^{2}\phi_{0}f(\phi_{0},0)\langle\hat{A}(\phi)\rangle_{\phi_{0}}$
which is particularly convenient for a initial Poisson distribution
of population, such that $\overline{A(N)}=\langle\hat{A}(\phi)\rangle_{\lambda}$,
where $\lambda$ is the initial mean.

\subsection{Noise-averaged pseudo-population\label{sub:Noise-averaged-pseudo-population}}

We now establish the statistical properties of the stochastic variable
$\phi(t)$. Our strategy will be to represent $\langle\phi\rangle_{\phi_{0}}$,
the average of Eq. (\ref{eq:119m}) over the noise for a given initial
value $\phi_{0}$, as a (formal) series in positive powers of $\phi_{0}$,
written
\begin{eqnarray}
\!\!\langle\phi\rangle_{\phi_{0}} & = & \sum_{j=1}^{\infty}C_{j}(t)\phi_{0}^{j}=\phi_{0}\sum_{n=0}^{\infty}(-1)^{n}(\kappa\phi_{0})^{n}M_{n}(t),\label{eq:128}
\end{eqnarray}
with $C_{j}(t)=(-\kappa)^{j-1}M_{j-1}(t)$ and where
\begin{equation}
M_{j}(t)=\left\langle e^{\kappa t+i(2\kappa)^{1/2}W_{t}}\left(\int_{0}^{t}e^{\kappa s+i(2\kappa)^{1/2}W_{s}}ds\right)^{j}\right\rangle .\label{eq:130}
\end{equation}
In Appendix \ref{sec:Formal-power-series} these coefficients are
evaluated. We find that $m_{n}(\kappa t)=[(2\kappa)^{n}/n!]M_{n}(t)$
satisfy the recursion relation
\begin{equation}
m_{n}=\sum_{l=1}^{n}\frac{2^{l}(2n-l)!(-1)^{l+1}}{(2n)!l!}m_{n-l},\label{eq:151}
\end{equation}
with
\begin{equation}
m_{0}(n)=1-\exp[-n(n+1)\kappa t],\label{eq:149}
\end{equation}
and thus all the $m_{n}$ may be generated, for example $m_{1}=1-\exp(-2\kappa t)$
and $m_{2}=\frac{1}{6}\left(2-3e^{-2\kappa t}+e^{-6\kappa t}\right)$,
and hence all the $C_{j}$ in Eq. (\ref{eq:128}). In Appendix \ref{sec:Mean-populations-in}
we explore how these coefficients, and the formal power series, allow
us to determine the mean and variance of the particle population as
a function of time and initial condition, recovering known results.

\subsection{Averaged pseudo-population as $t\to\infty$\label{sub:Averaged-pseudo-population-as}}

It will prove to be useful for comparisons with numerical work to
determine the value of $\langle\phi\rangle_{\phi_{0}}$ as $t\to\infty$.
Using Ito's lemma we have
\begin{equation}
d\phi^{n}=n\phi^{n-1}d\phi+\frac{1}{2}(-2\kappa\phi^{2})n(n-1)\phi^{n-2}dt,\label{eq:165}
\end{equation}
and we insert Eq. (\ref{eq:112}) and average such that
\begin{equation}
d\langle\phi^{n}\rangle=-\kappa n\langle\phi^{n+1}\rangle dt-\kappa n(n-1)\langle\phi^{n}\rangle dt.\label{eq:166}
\end{equation}
Assuming that as $t\rightarrow\infty$ all the $\langle\phi^{n}\rangle$
become time independent, we deduce that $\langle\phi^{n+1}\rangle=-(n-1)\langle\phi^{n}\rangle$
in this limit, in which case all moments for $n\ge2$ tend to zero,
for all initial conditions.

Next, we note that
\begin{equation}
d(e^{-\phi})=-e^{-\phi}d\phi-\kappa\phi^{2}e^{-\phi}dt=-i(2\kappa)\phi e^{-\phi}dW_{t},\label{eq:163}
\end{equation}
using Eq. (\ref{eq:112}), such that $d\langle\exp(-\phi)\rangle=0$,
and imposing the initial condition we obtain $\langle\exp(-\phi)\rangle_{\phi_{0}}=\exp(-\phi_{0})$
for all $t$. Since
\begin{equation}
\langle e^{-\phi}\rangle_{\phi_{0}}=\sum_{n=0}^{\infty}\frac{(-1)^{n}}{n!}\langle\phi^{n}\rangle_{\phi_{0}},\label{eq:167}
\end{equation}
we find that $\langle e^{-\phi}\rangle_{\phi_{0}}\to1-\langle\phi\rangle_{\phi_{0}}$
as $t\to\infty$, due to the vanishing of moments with $n\ge2$, and
we conclude that the stochastic dynamics of $\phi(t)$ give rise to
the asymptotic behaviour
\begin{equation}
\langle\phi\rangle_{\phi_{0}}\to1-\exp(-\phi_{0})\label{eq:167a}
\end{equation}
in the late time limit. We shall use this result to check the outcomes
of numerical calculations in  Section \ref{sec:Numerical-approach.}.
The result makes sense because the mean population arising from a
Poisson distribution with mean $\lambda$ at $t=0$ is equal to $\langle\phi\rangle_{\phi_{0}=\lambda}$.
Intuitively, all initial situations sampled from such a distribution
lead to a final population of unity as $t\to\infty$ except for the
case of an initial population of zero, the probability of which is
$\exp(-\lambda)$. An initial situation that leads to a final population
of unity is therefore selected with probability $1-\exp(-\lambda)$.
The mean population as $t\to\infty$ is $\overline{N}=0\times\exp(-\lambda)+1\times(1-\exp(-\lambda))=1-\exp(-\lambda)$.

We can confirm this result by considering kinetics starting from a
definite initial population $N_{0}$. From Eq. (\ref{eq:131}) we
write
\begin{equation}
\begin{aligned} & \overline{N}\to\oint_{C}d\phi_{0}\frac{N_{0}!}{2\pi i}\phi_{0}^{-N_{0}-1}\exp(\phi_{0})\left(1-\exp(-\phi_{0})\right)\\
 & =\oint_{C}d\phi_{0}\frac{N_{0}!}{2\pi i}\phi_{0}^{-N_{0}-1}\sum_{n=1}^{\infty}\frac{1}{n!}\phi_{0}^{n}\\
 & =\sum_{n=1}^{\infty}\delta_{N_{0}n}=1-\delta_{N_{0}0},
\end{aligned}
\label{eq:168}
\end{equation}
demonstrating that the asymptotic mean population is unity, unless
$N_{0}=0$ in which case it is zero.

\subsection{Source enhanced coagulation\label{sub:Source-enhanced-coagulation}}

We can also determine the asymptotic behaviour of a system of coagulating
species in which there is an injection rate $j$, to be used as a
check in later numerical calculations. The underlying master equation
is
\begin{equation}
\begin{aligned} & \frac{dP(N,t)}{dt}=jP(N-1,t)-jP(N,t)+\\
 & \kappa(N+1)NP(N+1,t)-\kappa N(N-1)P(N,t),
\end{aligned}
\label{eq:71}
\end{equation}
with the first term deleted for the case $N=0$. The evolution of
the corresponding pseudo-population is given by
\begin{equation}
d\phi=jdt-\kappa\phi^{2}dt+i(2\kappa)^{1/2}\phi dW_{t}.\label{eq:72}
\end{equation}
This time we find that
\begin{equation}
\begin{aligned} & d(e^{-\phi})=-e^{-\phi}d\phi-\kappa\phi^{2}e^{-\phi}dt\\
 & =-je^{-\phi}dt-i(2\kappa)^{1/2}\phi e^{-\phi}dW_{t},
\end{aligned}
\label{eq:73}
\end{equation}
so that $d\langle e^{-\phi}\rangle=-j\langle e^{-\phi}\rangle dt$
and $\langle e^{-\phi}\rangle\propto e^{-jt}$, with the proportionality
factor depending on the initial condition. We also have
\begin{equation}
d\langle\phi^{n}\rangle=n\langle\phi^{n-1}\left(j-\kappa\phi^{2}\right)\rangle dt-\kappa n(n-1)\langle\phi^{n}\rangle dt,\label{eq:75}
\end{equation}
which in a stationary state implies, for $n\ge1$, that
\begin{equation}
\langle\phi^{n+1}\rangle_{{\rm st}}=\frac{j}{\kappa}\langle\phi^{n-1}\rangle_{{\rm st}}-(n-1)\langle\phi^{n}\rangle_{{\rm st}}.\label{eq:76}
\end{equation}
This is reminiscent of a recurrence relation \cite{AbramStegun} for
modified Bessel functions, $I_{n-1}(x)-I_{n+1}(x)=(2n/x)I_{n}(x)$,
which suggests that $\langle\phi^{n}\rangle_{{\rm st}}\propto(j/\kappa)^{n/2}I_{n-1}(2(j/\kappa)^{1/2})$
for $n\ge0$. Since $\langle\phi^{2}\rangle_{{\rm st}}=j/\kappa$
, the proportionality factor is $\left[I_{1}(2(j/\kappa)^{1/2})\right]^{-1}$.
Notice that we do not label these stationary averages according to
$\phi_{0}$ since memory of the initial condition is lost for this
case. In the stationary state of a source enhanced coagulating system
we therefore expect the mean population to be
\begin{equation}
\overline{N}_{{\rm st}}=\langle\phi\rangle_{{\rm st}}=\left(\frac{j}{\kappa}\right)^{1/2}\frac{I_{0}\left(2(j/\kappa)^{1/2}\right)}{I_{1}\left(2(j/\kappa)^{1/2}\right)},\label{eq:77}
\end{equation}
which is similar to analysis performed elsewhere \cite{Gardiner09,Green01,Lushnikov03}.
The limit of this result as $j\to0$ does not correspond to Eq. (\ref{eq:167a})
since dependence on the initial condition reappears when $j=0$.

\subsection{Remarks}

We conclude our analytical consideration by noting that the coagulation
problem can be studied in a variety of ways, and that introducing
complex, stochastically evolving pseudo-populations, averaged over
the noise and over a complex contour of initial values (or over the
entire plane), might seem very elaborate compared with the direct
analytic solution to the master equations, for example. Our purpose,
however, is to establish explicitly that the approach works for a
simple case. Our attention now turns to the numerical solution of
the stochastic dynamics and averaging procedure to determine whether
this can be performed efficiently for the model. Establishing this
would suggest that a wider set of stochastic problems based on master
equations might be amenable to solution using these techniques.

\section{Numerical approach\label{sec:Numerical-approach.}}

\subsection{Stochastic numerics}

An analytical solution to the SDEs corresponding to more complicated
coagulation schemes, for example those modelled in the mean field
approximation using rate equations such as (\ref{eq:105}), is unlikely
to be available. In these cases our approach is implemented through
a numerical solution of the SDEs for the relevant pseudo-populations
$\phi_{i}(t)$, followed by averaging over the noise and a suitable
weighting of the results over the initial values $\phi_{i0}$. We
test the feasibility of this strategy for the simple $A+A\to A$ model.

\begin{figure}
\includegraphics[width=1\columnwidth]{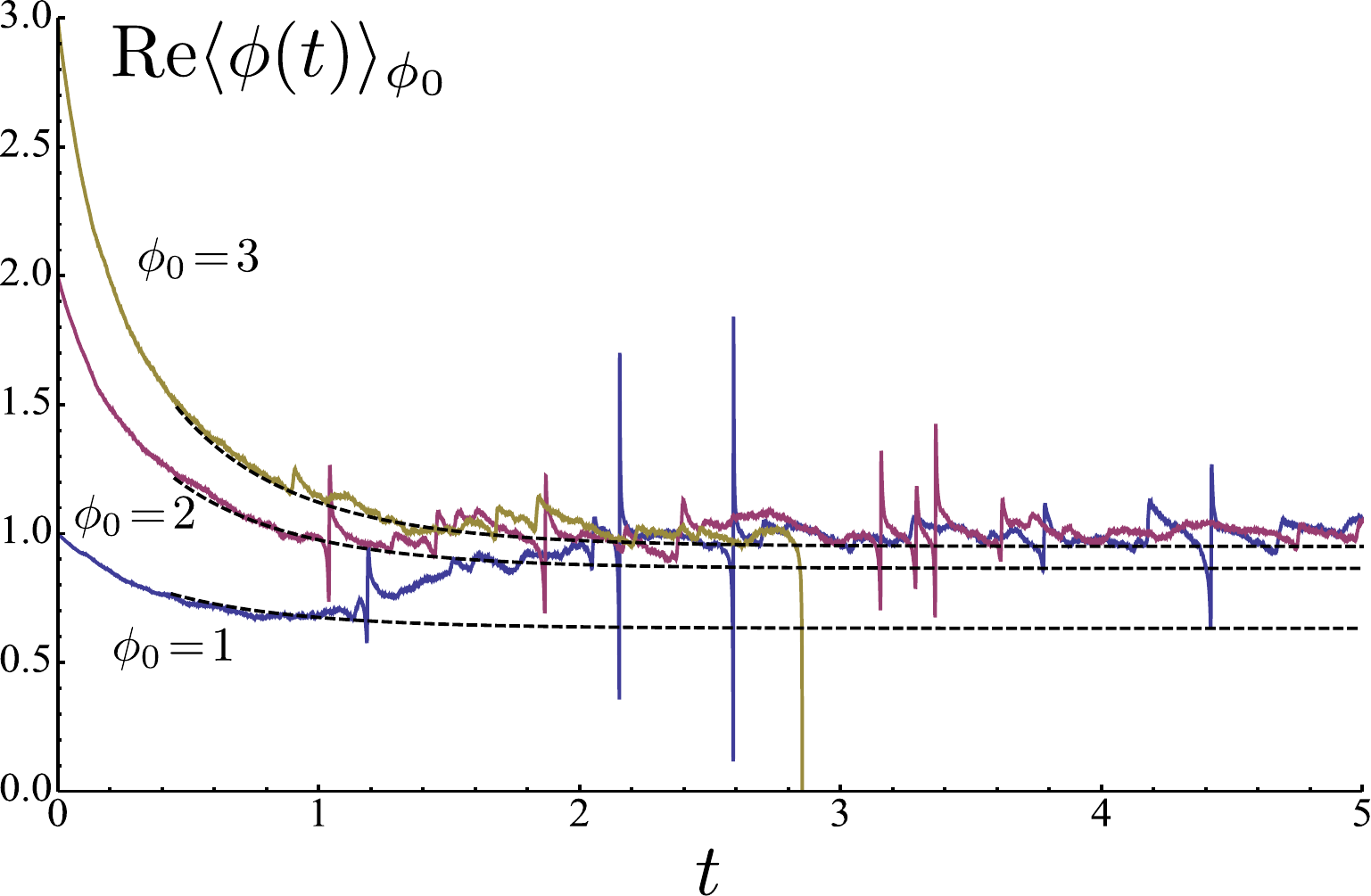}\protect\caption{Numerical solutions to SDE (\ref{eq:112}), averaged over $10^{3}$
realisations of the noise with a timestep of $5\times10^{-5}$ and
$\kappa=1$, and with initial values $\phi_{0}=1$, 2 and 3, compared
with series representations of $\langle\phi\rangle_{\phi_{0}}$ (dashed
lines) including terms up to 12th order in $\phi_{0}$, as derived
in Section \ref{sub:Noise-averaged-pseudo-population}. \label{fig:Solution-to-SDE}}
\end{figure}

We solve the SDE (\ref{eq:112}) numerically using a C++ code. In
Figure \ref{fig:Solution-to-SDE}, we plot $\mathrm{Re}\langle\phi\rangle_{\phi_{0}}$
against $t$ for $\phi_{0}=1$, 2 and 3, averaging $\phi$ over $10^{3}$
trajectories driven by independent realisations of the Wiener process
using $10^{5}$ timesteps of length $5\times10^{-5}$, and with $\kappa=1$.
Also shown are analytical results based on the formal series developed
in Section \ref{sub:Noise-averaged-pseudo-population} and Appendix
\ref{sec:Formal-power-series}, truncated at the 12th power of $\phi_{0}$.
This representation of the analytical solution appears to be accurate
for the chosen values of $\phi_{0}$. Asymptotically, the series approximates
well to values $1-\exp(-\phi_{0})$, in agreement with the analysis
in Section \ref{sub:Averaged-pseudo-population-as}. For small $t$
the outcomes resemble the mean field result Eq. (\ref{eq:102}), as
we suggest they should in Appendix \ref{sec:Steepest-descent-evaluation}.

However, the numerical results do not appear to be very satisfactory.
Firstly, the averages are noisy, suggesting that $10^{3}$ realisations
are too few in number to obtain statistical accuracy, though this
can of course be increased. Secondly, they do not seem to possess
the correct limits at large times: they all seem to tend towards unity.
Thirdly, they often exhibit sharp peaks, which are sometimes large
enough to cause the numerical simulation to fail, as is seen in the
case for $\phi_{0}=3$, where a negative spike at $t\approx2.85$
causes the computation to crash. These instabilities, even if not
terminal, have a disproportionate effect on the statistics when the
number of realisations is limited. Such instabilities have been encountered
before in simulations \cite{Drummond04,Gardiner09}, and in the next
section we examine their origin and consider how they can be avoided.

\subsection{Elimination of instabilities\label{sub:Elimination-of-instabilities}}

\begin{figure}
\begin{centering}
\includegraphics[width=1\columnwidth]{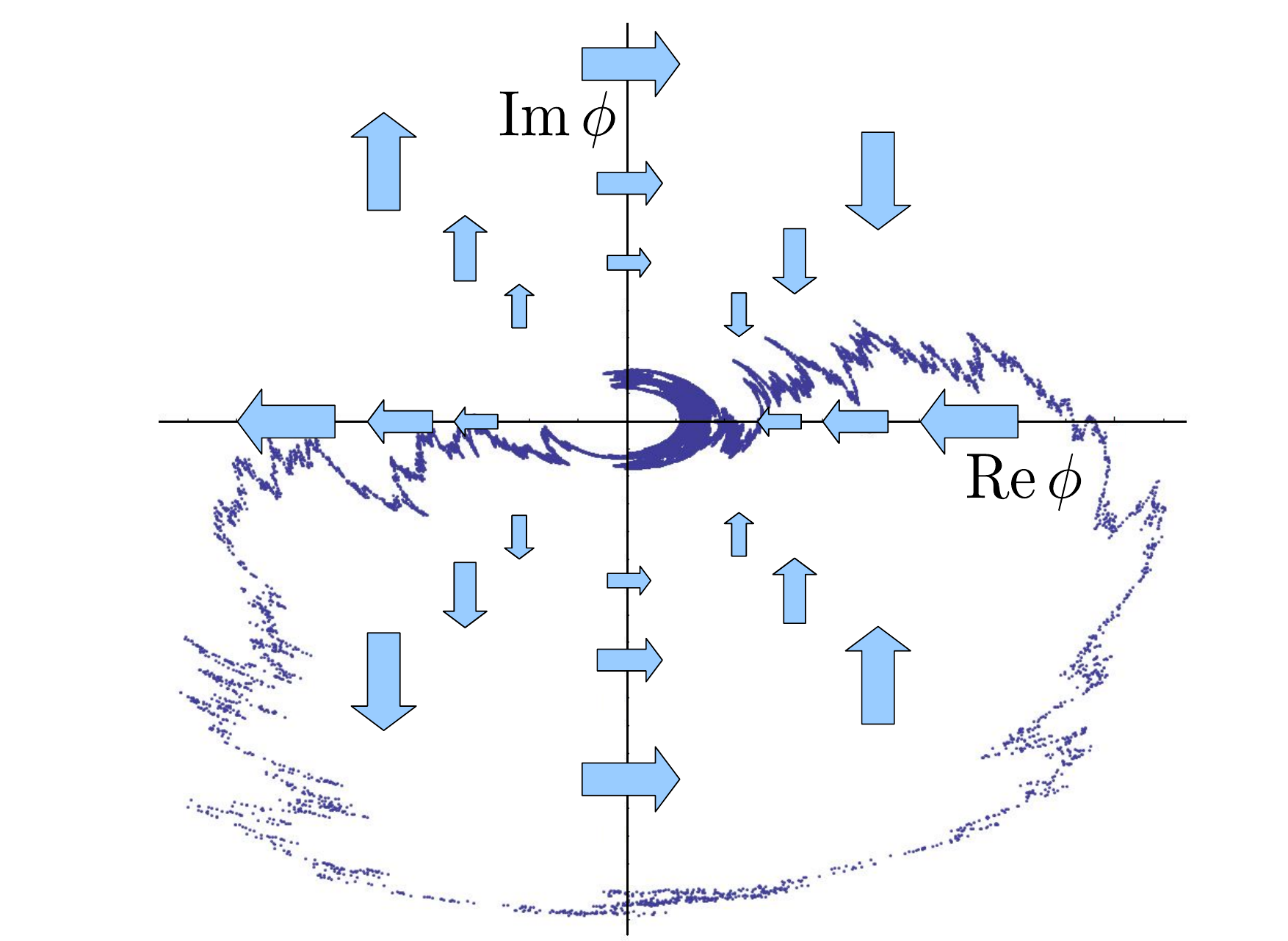}
\par\end{centering}

\protect\caption{The block arrows indicate the magnitude and direction of drift of
$\phi(t)$ across the complex plane brought about by the deterministic
term $-\kappa\phi^{2}dt$ in the SDE (\ref{eq:112}). The effect of
this pattern is seen in an example realisation of the evolution. An
excursion towards negative Re$(\phi)$ from the quasistable crescent
shaped region near the origin, followed by a return path taken in
an anticlockwise direction, is responsible for a spike in the average
of $\phi$ shown in Figure \ref{fig:Solution-to-SDE}. \label{fig:Drift-directions-over} }
\end{figure}

The origin of the large fluctuations in $\phi(t)$ can be understood
by considering the deterministic contribution to its evolution in
the complex plane according to Eq. (\ref{eq:112}). This is shown
schematically by block arrows in Figure \ref{fig:Drift-directions-over},
representing the magnitude and orientation of the complex quantity
$-\phi^{2}$. Were it not for stochastic noise, a trajectory that
started with a positive value of Re$\phi$ would remain in the right
hand side of the complex plane and indeed would drift asymptotically
towards the origin. The picture is quite different for a negative
initial value of Re$\phi$: there is a component of drift away from
the origin, though in most cases the path ultimately moves around
into the right hand side of the plane and from there to the origin.
The exception is for trajectories starting on the negative real axis,
which asymptotically move towards $-\infty$. This pattern of drift
has the potential for instability.

Of course the evolution of $\phi$ is also driven by noise, and an
example trajectory is shown in Figure \ref{fig:Drift-directions-over}.
For much of the time $\phi$ evolves in a rather well-behaved fashion
over the complex plane, forming a crescent shaped trace. The noise
maintains the trajectory away from the deterministic attractor at
the origin. However, noise is also the undoing of this stability:
if the trajectory wanders onto the left hand side of the plane driven
by the stochastic term in Eq. (\ref{eq:112}), then a tendency to
drift towards $\phi=-\infty$ can set in. This is also shown in Figure
\ref{fig:Drift-directions-over}: at some point a fluctuation causes
$\phi(t)$ to move towards the left, away from the crescent pattern.
In this example, quite a significant excursion towards negative real
values occurs before noise deflects the trajectory sufficiently away
from the real axis, allowing the underlying drift to take $\phi(t)$
in an anticlockwise fashion back towards the right hand side of the
plane. Once there, the $\phi(t)$ settles again into the crescent
pattern. The intervals spent within the crescent correspond to those
periods in Figure \ref{fig:Solution-to-SDE} where fluctuations are
small, whilst excursions into the left hand plane seem to be responsible
for the spikes \cite{Deloubriere02}.

These instabilities are not desirable and action should be taken to
eliminate them. They are less common if the timestep in the numerical
simulation is reduced, since this lowers the likelihood of a noise-driven
jump into the left hand side of the complex plane, but this increases
the computational expense of the approach. We therefore need a mathematical
scheme that can suppress the dangerous drift pattern in the stochastic
dynamics while retaining the statistical properties of solutions to
the SDE.

Drummond \cite{Drummond04} proposed a scheme for the elimination
of such instabilities, taking the form of a modification to the fundamental
Poisson representation through the introduction of a weighting parameter
$\Omega(t)$, chosen to evolve according to
\begin{equation}
d\Omega=\Omega gdW_{t},\label{eq:169}
\end{equation}
with initial condition $\Omega(0)=1$, and where $g$ is an arbitrary
function. A new variable $\phi^{\prime}$ is introduced, subject to
the same initial condition imposed on $\phi$, but evolving according
to
\begin{equation}
d\phi^{\prime}=-\kappa\phi^{\prime2}dt+i(2\kappa)^{1/2}\phi^{\prime}\left(dW_{t}-gdt\right),\label{eq:170}
\end{equation}
and Drummond showed that the $\Omega$-weighted average of $\phi^{\prime}$
over the noise is the same as the corresponding average of $\phi$:
\begin{equation}
\langle\phi\rangle=\langle\Omega\phi^{\prime}\rangle.\label{eq:171}
\end{equation}
However, since $\phi^{\prime}$ and $\phi$ evolve according to different
SDEs, they need not suffer from the same instabilities.

Drummond termed this the `gauging away'\emph{ }of the instabilities
of the original SDE, the terminology suggested by recognising that
Eq. (\ref{eq:171}) possesses an invariance with respect to different
choices of $g$. The evolution of $\phi^{\prime}$ and $\Omega$ is
affected by the form of $g,$ but not $\langle\phi\rangle$. In other
areas of theoretical physics, especially in electromagnetism and quantum
field theories, a gauge transformation is precisely a recasting of
a theoretical problem that has no effect on the eventual physical
predictions, and hence the terminology is appropriate. The transformation
can be useful mathematically but is arbitrary as far as the physics
is concerned.

The approach is in fact equivalent to a transformation of the probability
measure in stochastic calculus, and can be perhaps be most easily
understood as an application of the Cameron-Martin-Girsanov formula
\cite{CameronMartin44,CameronMartin45,Girsanov60} in stochastic calculus,
as we describe in more detail in Appendix \ref{sec:Equivalence-of-Drummond}.

We explore one of Drummond's gauge functions $g(\phi^{\prime})$ that
has the capacity to tame the instability in the coagulation problem.
Consider the choice
\begin{equation}
g=i\left(\kappa/2\right)^{1/2}\left(\phi^{\prime}-\vert\phi^{\prime}\vert\right),\label{eq:172}
\end{equation}
which leads to the SDEs
\begin{equation}
d\phi^{\prime}=-\kappa\phi^{\prime}\vert\phi^{\prime}\vert dt+i(2\kappa)^{1/2}\phi^{\prime}dW_{t},\label{eq:173}
\end{equation}
 and
\begin{equation}
d\Omega=i\left(\kappa/2\right)^{1/2}\left(\phi^{\prime}-\vert\phi^{\prime}\vert\right)\Omega dW_{t}.\label{eq:174}
\end{equation}

The crucial difference between Eqs. (\ref{eq:173}) and (\ref{eq:112})
is that the drift term in the SDE for $\phi^{\prime}$ is always directed
towards the origin. It lacks the ability to produce an excursion towards
$\phi^{\prime}=-\infty$. Meanwhile, $\Omega(t)$ evolves diffusively
in the complex plane, with no drift and hence $d\langle\Omega\rangle=0$.
As long as $\phi^{\prime}$ is real, the increment $d\Omega$ vanishes,
and because of this feature, Drummond called this choice of $g$ a
\emph{minimal} gauge function, and regarded it as a natural choice
for cases where $\phi^{\prime}$ is expected to take real values for
most of its history.

We have studied this reformulation of the coagulation problem and
confirmed numerically that $\phi^{\prime}(t)$ does not suffer from
instabilities of the kind experienced by $\phi(t)$, and that the
$\Omega$-weighted average of $\phi^{\prime}$ appears to agree with
various analytic results expected for $\langle\phi\rangle_{\phi_{0}}$,
as shown in Figure \ref{fig:-against-time,}. Nevertheless, we notice
that the statistical uncertainty in the average of $\Omega(t)$ grows
as time progresses, and this introduces a decline in accuracy, for
a given number of realisations. We next address the reasons for this.

\begin{figure}
\includegraphics[width=1\columnwidth]{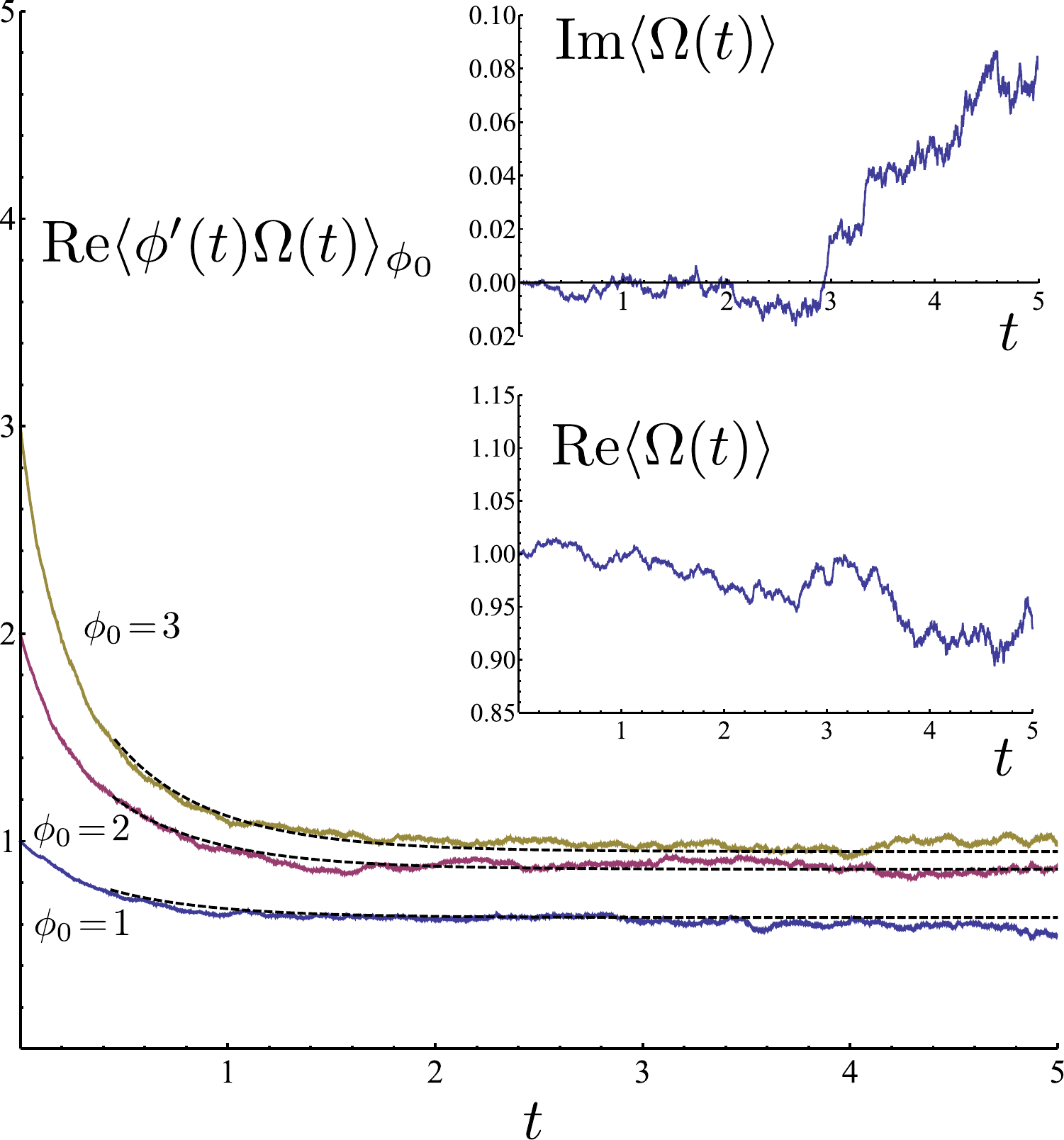}\protect\caption{Evolution of Re$\langle\phi'\Omega\rangle_{\phi_{0}}$ according to
Eqs. (\ref{eq:173}) and (\ref{eq:174}), with $\kappa=1$ and $\phi_{0}=1$,
2 and 3, for the same timestep and number of realisations as in Figure
\ref{fig:Solution-to-SDE}. The expected analytic behaviour is indicated,
and in comparison with Figure \ref{fig:Solution-to-SDE} the noise
in the numerical results is much reduced, which is a consequence of
the gauging procedure. The evolution of the real and imaginary parts
of $\langle\Omega\rangle$ for the case with $\phi_{0}=2$ is shown
in the inset, to illustrate the deviation from the expected values
of unity and zero, respectively, as $t$ increases: this is caused
by sampling errors. \label{fig:-against-time,}}
\end{figure}

\subsection{Asymptotic behaviour of $\Omega$}

The diffusive behaviour of the weight function $\Omega$ can be best
demonstrated by considering the evolution of the mean of $\vert\Omega\vert^{2}$.
We have $d\Omega=g\Omega dW_{t}$ and hence $d\Omega^{*}=g^{*}\Omega^{*}dW_{t}$
such that, using Ito's lemma
\begin{equation}
\begin{aligned} & d\vert\Omega\vert^{2}=\Omega d\Omega^{*}+\Omega^{*}d\Omega+g\Omega g^{*}\Omega^{*}dt\\
 & =\Omega g^{*}\Omega^{*}dW_{t}+\Omega^{*}g\Omega dW_{t}+\vert g\vert^{2}\vert\Omega\vert^{2}dt,
\end{aligned}
\label{eq:175}
\end{equation}
 and so
\begin{equation}
d\langle\vert\Omega\vert^{2}\rangle=\langle\vert g\vert^{2}\vert\Omega\vert^{2}\rangle dt,\label{eq:176}
\end{equation}
which indicates that the mean square modulus of $\Omega$ increases
monotonically, whatever choice of function $g$ is made. The mean
of $\Omega$ is unity for all $t$, but its increasing mean square
modulus suggests that the distribution of $\Omega$ broadens as $t$
increases. As a consequence, the accurate extraction of the statistical
properties of the dynamics will require more realisations for larger
$t$. This is the price to pay, it seems, for the taming of the instabilities
in the original SDE.

\subsection{Optimised gauge}

Nevertheless, the choice available to us in the gauging procedure,
or equivalently the freedom to shift the probability measure according
to the Cameron-Martin-Girsanov formula (see Appendix \ref{sec:Equivalence-of-Drummond}),
allows us to optimise the quality of the numerical estimates of $\langle\phi\rangle$.
Consider a new gauge function containing the real variable $R$:
\begin{equation}
g=i\left(\kappa/2\right)^{1/2}\left(\phi^{\prime}-\vert\phi^{\prime}\vert-R\right),\label{eq:177}
\end{equation}
which differs from Eq. (\ref{eq:172}) by an imaginary constant. It
leads to the SDEs
\begin{equation}
d\phi^{\prime}=-\kappa\phi^{\prime}\left(R+\vert\phi^{\prime}\vert\right)dt+i(2\kappa)^{1/2}\phi^{\prime}dW_{t},\label{eq:178}
\end{equation}
and
\begin{equation}
d\Omega=i\left(\kappa/2\right)^{1/2}\left(\phi^{\prime}-\vert\phi^{\prime}\vert-R\right)\Omega dW_{t},\label{eq:179}
\end{equation}
with the property $\langle\phi\rangle=\langle\Omega\phi^{\prime}\rangle$.
Intuitively, a positive $R$ has the effect of strengthening the drift
towards the origin in the SDE for $\phi^{\prime}$. It is revealing
to study the evolution of the square modulus of $\phi^{\prime}$ under
this gauge. We write
\begin{align}
 & d(\phi^{\prime}\phi^{\prime*})=\phi^{\prime}d\phi^{\prime*}+\phi^{\prime*}d\phi^{\prime}+i(2\kappa)^{1/2}\phi^{\prime}\left[-i(2\kappa)^{1/2}\phi^{\prime*}\right]dt\nonumber \\
 & =-\kappa\vert\phi^{\prime}\vert^{2}\left(R+\vert\phi^{\prime}\vert\right)dt-i(2\kappa)^{1/2}\vert\phi^{\prime}\vert^{2}dW_{t}\nonumber \\
 & \quad-\kappa\vert\phi^{\prime}\vert^{2}\left(R+\vert\phi^{\prime}\vert\right)dt+i(2\kappa)^{1/2}\vert\phi^{\prime}\vert^{2}dW_{t}+2\kappa\vert\phi^{\prime}\vert^{2}dt,\label{eq:179a}
\end{align}
so
\begin{equation}
d\vert\phi^{\prime}\vert^{2}=2\kappa(1-R)\vert\phi^{\prime}\vert^{2}-2\kappa\vert\phi^{\prime}\vert^{3},\label{eq:179b}
\end{equation}
and thus $\vert\phi^{\prime}\vert$ evolves deterministically towards
an asymptotic value of $(1-R)$, if $R\le1$, or zero if $R>1$.

It is also revealing to consider the SDE for $\Omega\phi^{\prime}$
explicitly:
\begin{align}
 & d(\Omega\phi^{\prime})=\phi^{\prime}d\Omega+\Omega d\phi^{\prime}\nonumber \\
 & \quad+i\left(\kappa/2\right)^{1/2}\left(\phi^{\prime}-\vert\phi^{\prime}\vert-R\right)\Omega i(2\kappa)^{1/2}\phi^{\prime}dt\nonumber \\
 & =\phi^{\prime}i\left(\kappa/2\right)^{1/2}\left(\phi^{\prime}-\vert\phi^{\prime}\vert-R\right)\Omega dW_{t}\nonumber \\
 & \quad+\Omega\left(-\kappa\phi^{\prime}\left(R+\vert\phi^{\prime}\vert\right)dt+i(2\kappa)^{1/2}\phi^{\prime}dW_{t}\right)\nonumber \\
 & \quad-\kappa\phi^{\prime}\Omega\left(\phi^{\prime}-\vert\phi^{\prime}\vert-R\right)dt\nonumber \\
 & =-\kappa\Omega\phi^{\prime2}dt+i\left(\kappa/2\right)^{1/2}\phi^{\prime}\left(\phi^{\prime}-\vert\phi^{\prime}\vert-R+2\right)\Omega dW_{t},\label{eq:179c}
\end{align}
such that as $t\to\infty$ the stochastic term on the right hand side
tends towards either $i\left(\kappa/2\right)^{1/2}\phi^{\prime}\left((1-R)\exp[i{\rm \, arg}(\phi^{\prime})]+1\right)\Omega dW_{t}$
if $R\le1$, or $i\left(\kappa/2\right)^{1/2}\phi^{\prime}\left(2-R\right)\Omega dW_{t}$
if $R>1$. Furthermore, the evolution of the average square modulus
of $\Omega$ is described by
\begin{equation}
\begin{aligned} & d\langle\vert\Omega\vert^{2}\rangle=\frac{\kappa}{2}\langle\vert\phi^{\prime}-\vert\phi^{\prime}\vert-R\vert^{2}\vert\Omega\vert^{2}\rangle dt\end{aligned}
,\label{eq:179d}
\end{equation}
which tends to $\frac{1}{2}\kappa\langle\vert(1-R)\exp[i\,{\rm arg}(\phi^{\prime})]-1\vert^{2}\vert\Omega\vert^{2}\rangle dt$
if $R\le1$, or $\frac{1}{2}\kappa R{}^{2}\langle\vert\Omega\vert^{2}\rangle dt$
if $R>1$.

This analysis indicates that there are advantages in making specific
choices of $R$. If we choose $R=2$, Eq. (\ref{eq:179c}) shows that
we can eliminate the noise term controlling the asymptotic time evolution
of $\Omega\phi^{\prime}$. On the other hand, it is the diffusive
behaviour of $\Omega$ that characterises the asymptotic statistics
of other quantities. The strength of the diffusion depends on the
quantity $(1-R)^{2}-2(1-R)\cos{\rm arg}(\phi^{\prime})+1$ for $R\le1$,
and $R^{2}$ for $R>1$. It is clear that the asymptotic rate of diffusive
spread of $\Omega$ increases as $R$ becomes large, which is disadvantageous.

\begin{figure}
\includegraphics[width=1\columnwidth]{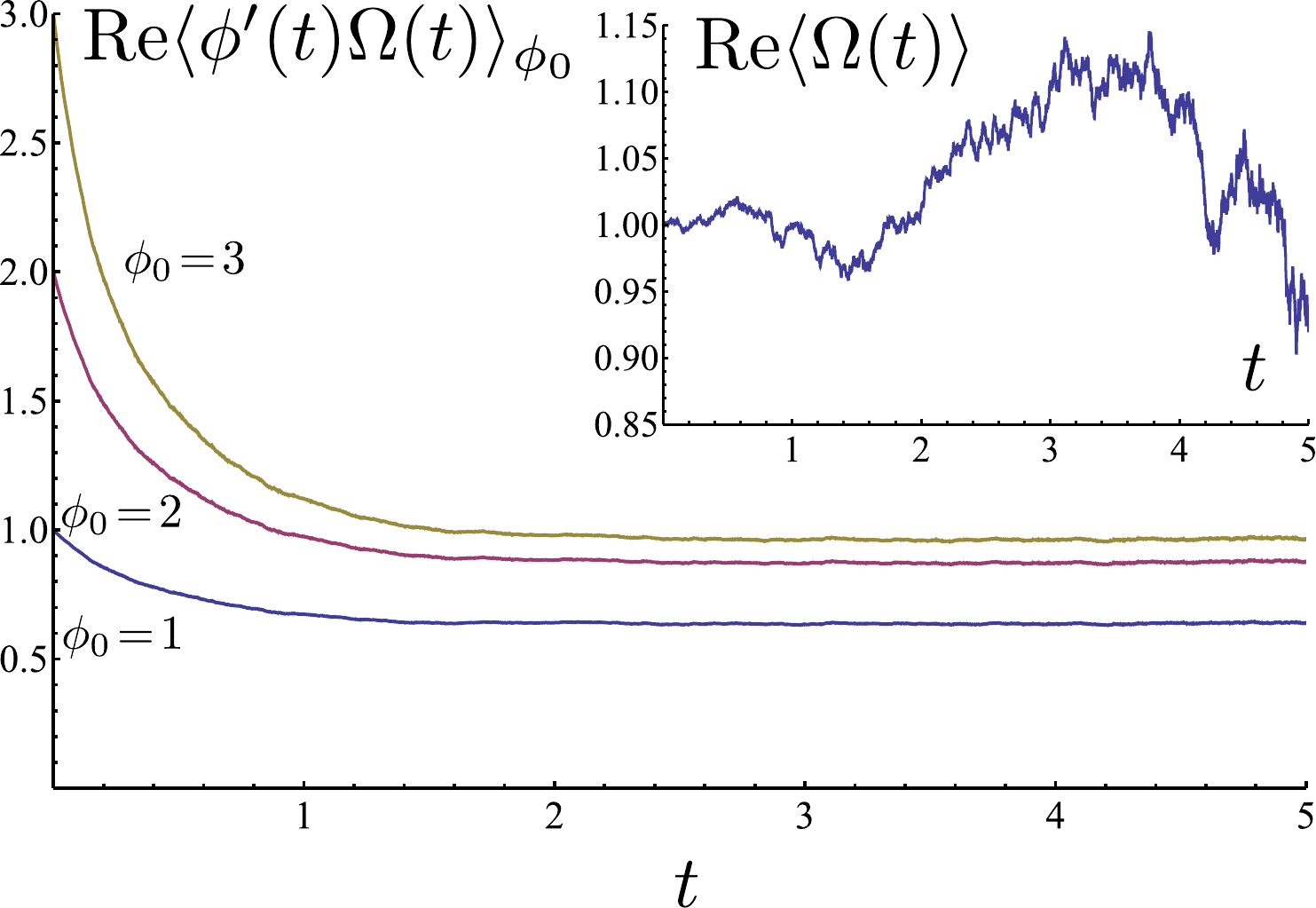}

\protect\caption{Evolution of Re$\langle\Omega\phi^{\prime}\rangle_{\phi_{0}}$ gauged
according to the improved procedure described by Eqs. (\ref{eq:178})
and (\ref{eq:179}) with $R=1$, for initial conditions $\phi_{0}=1$,
2 and 3. This should be compared with the results from Eqs. (\ref{eq:173})
and (\ref{eq:174}) in Figure \ref{fig:-against-time,}, obtained
with the same timestep and number of realisations but using $R=0$.
The evolution is less affected by statistical noise and corresponds
very closely to the expected analytic behaviour. The evolution of
$\Omega$ is more noisy, however, as indicated in the inset. \label{fig:efficacy}}

\end{figure}

As a consequence, we have explored gauges specified by values of $R$
between zero and two. As $R$ increases, the diffusive behaviour of
$\Omega$ becomes more marked while the noise in the evolution of
$\Omega\phi^{\prime}$ reduces. We find that the choice $R=1$ strikes
a balance by suppressing noise in the evolution of ${\rm Re}\langle\phi\rangle_{\phi_{0}}={\rm Re}\langle\Omega\phi^{\prime}\rangle_{\phi_{0}}$,
at the price of more significant statistical deviations of Re$\langle\Omega\rangle$
from the expected value of unity than in the case $R=0$. This is
illustrated in Figure \ref{fig:efficacy} which is to be compared
with Figure \ref{fig:-against-time,}. The shift in gauge seems to
have transferred the statistical noise from $\langle\Omega\phi^{\prime}\rangle$
into $\langle\Omega\rangle$. We suggest that $R=1$ specifies an
optimised gauge for this stochastic problem.

\subsection{Further calculations}

\begin{figure}
\includegraphics[width=1\columnwidth]{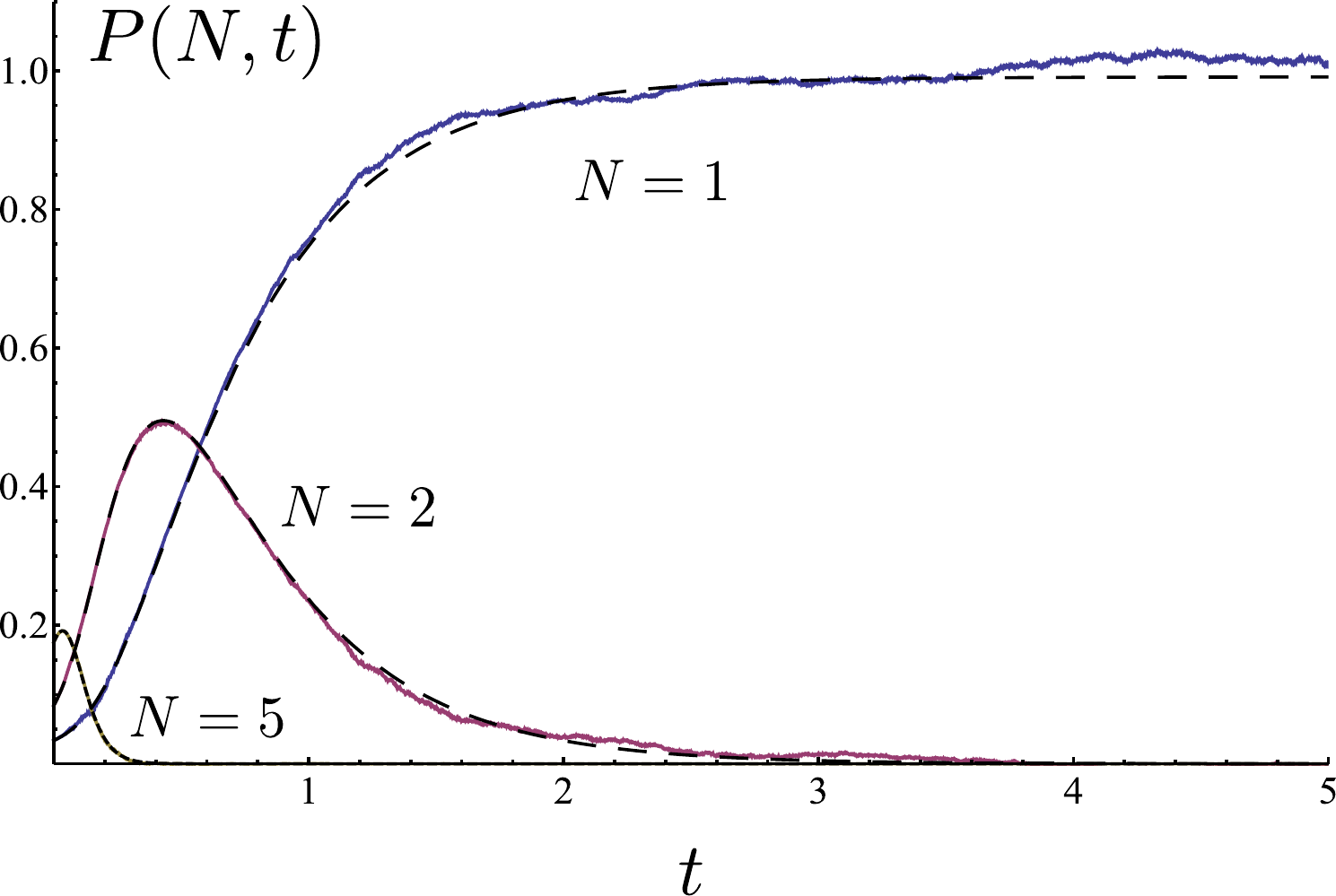}

\protect\caption{Probabilities of the occurrence of five, two and one particle as coagulation
progresses for $\kappa=1$, starting from a Poisson distribution with
mean $\lambda=5$ at $t=0$, comparing numerical evaluations of Eq.
(\ref{eq:78}), gauged through Eq. (\ref{eq:177}) with $R=1$ (solid
lines) with solutions to the master equations (\ref{eq:100}) (dashed
lines).\label{fig:probabilities.}}

\end{figure}

Using the gauge function (\ref{eq:177}) with $R=1$ we can study
further aspects of the pseudo-population dynamics that were investigated
analytically in Section \ref{sec:Analytic-model-of}. For particle
coagulation starting from a Poisson distribution with mean $\lambda$
the averages are simply given by quantities $\overline{A(N)}=\langle\hat{A}(\phi)\rangle_{\lambda}$.
In Figure \ref{fig:probabilities.} we plot the evolution of a selection
of state probabilities $P(N,t)$ for $\lambda=5$, derived from Eq.
(\ref{eq:78}), together with the solutions to Eq. (\ref{eq:100})
for a similar case with the initial Poisson distribution truncated
at $N=12$ for convenience. The timestep and number of realisations
are the same as in earlier cases. The correspondence is very good,
and the noise does not significantly distort the results over the
time scale for the completion of the coagulation.

\begin{figure}
\includegraphics[width=1\columnwidth]{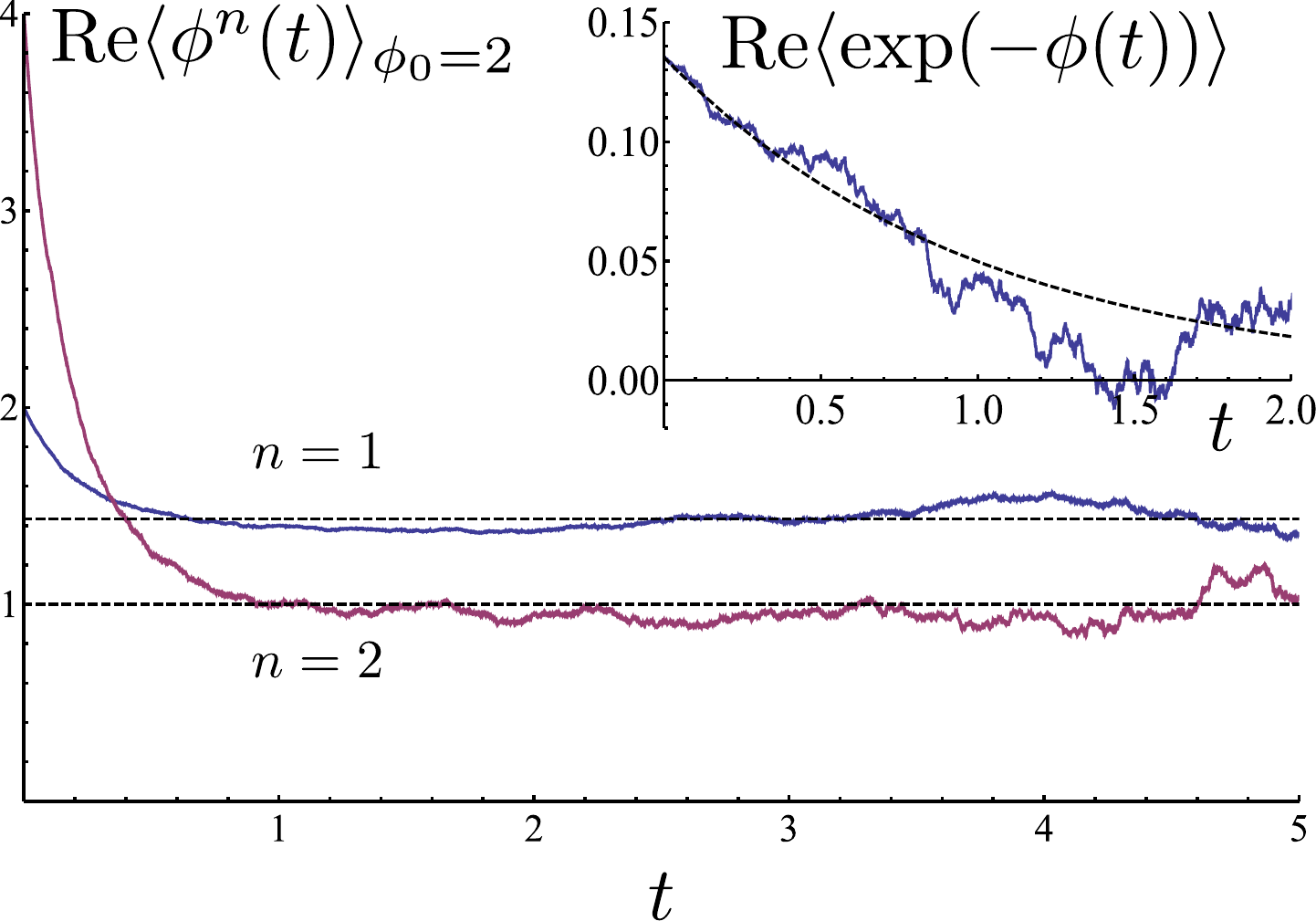}

\protect\caption{Main image: evolution of $\langle\phi\rangle_{\phi_{0}}$ and $\langle\phi^{2}\rangle_{\phi_{0}}$
(solid lines) according to the source enhanced coagulation kinetics
embodied in Eq. (\ref{eq:72}), with $j=\kappa=1$, initial condition
$\phi_{0}=\lambda=2$, and $10^{3}$ realisations with time step $5\times10^{-5}$.
At late times they correspond well to the stationary values $\langle\phi\rangle_{{\rm st}}=I_{0}(2)/I_{1}(2)\approx1.43$
and $\langle\phi^{2}\rangle_{{\rm st}}=1$ (dashed lines) derived
in Section \ref{sub:Source-enhanced-coagulation}. Inset: a numerical
calculation of the real part of $\langle\exp(-\phi)\rangle$ (solid
line) is compared with the expected behaviour $\exp(-2-jt$) (dashed
line), to illustrate that not every variable has acceptable statistics
for this number of realisations of the stochastic evolution. \label{fig:source-enhanced}}

\end{figure}

Finally, we examine source enhanced coagulation modelled by Eq. (\ref{eq:72}),
with the use of the optimised gauge. The quantities $\langle\phi\rangle=\langle\Omega\phi^{\prime}\rangle$
and $\langle\phi^{2}\rangle=\langle\Omega\phi^{\prime2}\rangle$ are
plotted for $j=\kappa=1$ in Figure \ref{fig:source-enhanced} to
show they compare well with the analytical results for late times
obtained in Section \ref{sub:Source-enhanced-coagulation}. The initial
condition is a Poisson distribution with $\lambda=2$. However, to
indicate that not all quantities are statistically well represented,
we also show the average of $\exp(-\phi)$, which ought to equal $\exp(-2-jt$)
for this case. Clearly more realisations would be needed to reproduce
this behaviour accurately with this choice of gauge.

\section{\label{sec:Discussion}Discussion}

We have outlined in this paper how to solve a simple coagulation problem
using some rather elaborate analytical and numerical methods, retrieving
results that were previously either known, or computable from the
master equations. The effort has not been misdirected, however, since
our aim has been to establish that the approach can be implemented
successfully for a well characterised example problem. The greater
part of our interest lies in more complicated cases of coagulation
such as those with multiple species that agglomerate at a variety
of rates. In such systems we expect to find behaviour that arises
due to statistical fluctuations around low mean populations that cannot
be captured in the usual mean field formulation. We plan to address
such cases in further work.

Specifically, we employ a Poisson representation \cite{Gardiner77,Gardiner09}
of the probability $P(N,t)$ that there should be $N$ particles present
at time $t$ in the simple case of $A+A\to A$ kinetics. The Poisson
representation is a superposition of Poisson distributions with complex
means. The description can be cast as a problem in the stochastic
dynamics of a complex pseudo-population, the average of which over
the noise and the initial condition is related to the average of the
physical particle population. Analytical work gives a series expansion
of this average in powers of the initial value of the pseudo-population,
and this can be employed to recover known results \cite{Green01,Deloubriere02,Lushnikov03,Tauber05,Barzykin05,Hochberg06}.
The development of the series involves the evaluation of multiple
integrals of functions of the Wiener process. We have also identified
some exact results valid in the limit $t\to\infty$, and in Appendix
\ref{sec:Steepest-descent-evaluation} give an approximate result
for small $\kappa t$ that resembles a mean field solution to the
kinetics. We have discussed the evaluation of averages of general
functions of the population, such as higher moments. We have also
studied the stationary state of a coagulating system enhanced by a
constant injection of new particles.

Analytical work can only be performed in certain circumstances, and
more generally a numerical solution of the stochastic differential
equations (SDEs) is necessary, followed by averaging over noise and
initial condition. Unfortunately, numerical instabilities make this
problematic, as has been noted previously \cite{Deloubriere02,Drummond04,Gardiner09}.
However, we follow Drummond \cite{Drummond04} in identifying an SDE
for a pseudo-population that is free of instabilities, to be used
together with a weighting procedure that reproduces the statistics
of the desired system. In Appendix \ref{sec:Equivalence-of-Drummond}
we have shown that this `gauging' scheme is equivalent to an adaptive
shift in the probability measure for the stochastic variable in the
SDE, and that the weighting procedure is an application of the Cameron-Martin-Girsanov
formula. This interpretation arguably makes the gauging procedure
more intuitive.

We have identified a choice of gauge, and associated transformation
of the problem, that provides statistical information in a rather
optimised fashion. With this approach, the growth with time in the
statistical uncertainty that is inherent with gauging can be adequately
controlled, such that we need to generate relatively few realisations
of the evolution, requiring rather little computational effort. A
number of the analytically derived results have been reproduced using
this numerical approach.

Of course Monte Carlo (MC) simulation of coagulation events taking
place in an evolving population of particles would be an obvious alternative
numerical method for studying this system. It is readily implemented,
for example using the Gillespie algorithm \cite{Gillespie77}, and
has the capacity to include population fluctuation effects. It is
an approach that is probably easier to grasp than the methods we have
outlined, and more computationally efficient for the problem under
consideration here. Nevertheless, we believe that pseudo-population
methods will prove to be more efficient than MC in coagulation problems
involving multiple species.

This point can be made by counting the number of master equations
that would effectively be solved by an MC procedure. If we wish to
study the evolution of clusters of monomers that have size dependent
agglomeration properties, then each size must be treated as a distinct
species. If we consider clusters ranging in size from 1 to $N_{{\rm max}}$
monomers, then master equations describing the evolution of probabilities
$P(N_{1},N_{2},\cdots,N_{N_{{\rm max}}},t)$ must be solved, where
$N_{i}$ denotes the population of clusters of size $i$. If each
population can vary between 0 and $n_{i}^{{\rm max}}$, then the number
of elementary population distributions that must be considered is
of order $\prod_{i=1}^{N_{{\rm max}}}\left(n_{i}^{{\rm max}}+1\right)$.
It is reasonable to say that this number can become rather large and
will grow faster than $N_{{\rm max}}$. The solution to such a large
number of coupled ordinary differential equations, or the equivalent MC simulation, would be
daunting.

On the other hand, the stochastic approach requires the solution to
just $N_{{\rm max}}$ SDEs for pseudo-populations $\phi_{i}(t)$;
these are analogues of the mean populations of clusters of size $i$.
The SDEs would bear some resemblance to the Smoluchowski coagulation
equations for mean populations in the mean field, fluctuation-free
limit, given in Eq. (\ref{eq:105}). Analytic solution to these equations
might be unavailable, but numerical solution would not be difficult,
especially if we have techniques such as gauging at our disposal to
avoid some of the pitfalls we have identified. The drawback is that
averaging of the SDEs over a variety of noise histories and perhaps
initial conditions is necessary. Nevertheless, the task is linear
in $N_{{\rm max}}$ and must eventually become more efficient than
direct solution to the master equations as $N_{{\rm max}}$ increases.

It is only when the mean of a population becomes small that deviations
from mean field behaviour emerge, and so a hybrid approach might be
possible whereby mean field rate equations are used to model the early
stages of coagulation, going over to pseudo-population rate equations
when the mean population becomes small. This is possible since the
mean field rate equations closely resemble the evolution equations
of the pseudo-populations, which is an important conceptual and mathematical
point of correspondence. Specifically, we could integrate the equations
for the $\phi_{i}(t)$ with the neglect of the noise when the modulus
of $\phi_{i}(t)$ is large, such that they remain real, and introduce
noise, and average over it, only when the modulus becomes small.

In conclusion, we have made conceptual and numerical developments
of a method for kinetic modelling that was introduced some time ago
\cite{Gardiner77} but that appears not to have been fully exploited.
We believe that in spite of some complexity in the formulation, the
approach possesses considerable intuitive value, and that it has the
capability to treat systems with interesting statistical properties
for which alternative methods are inappropriate or expensive. We intend
to explore these possibilities in further studies.
\begin{acknowledgments}
Funding from the Leverhulme Trust through grant F/07/134/BV is gratefully
acknowledged.
\end{acknowledgments}

\appendix

\section{Stochastic evolution of a complex pseudo-population \label{sec:Stochastic-evolution-of}}

We solve Eq. (\ref{eq:112}) to justify the expression for $\phi(t)$
given in Eq. (\ref{eq:119m}). Consider $G(\phi,x)=\phi^{-1}\exp(x)$
with stochastic variables evolving according to $d\phi=adt+bdW_{t}$
and $dx=pdt+qdW_{t}$. By Ito's lemma we have
\begin{equation}
dG=\frac{\partial G}{\partial\phi}d\phi+\frac{\partial G}{\partial x}dx+\frac{1}{2}b^{2}\frac{\partial^{2}G}{\partial\phi^{2}}dt+bq\frac{\partial^{2}G}{\partial\phi\partial x}dt+\frac{1}{2}q^{2}\frac{\partial^{2}G}{\partial x^{2}}dt,\label{eq:113}
\end{equation}
so
\begin{align}
 & dG=-\frac{1}{\phi}Gd\phi+Gdx+\frac{b^{2}}{\phi^{2}}Gdt-\frac{bq}{\phi}Gdt+\frac{q^{2}}{2}Gdt\label{eq:114}\\
 & =\left(-\frac{a}{\phi}+p+\frac{b^{2}}{\phi^{2}}-\frac{bq}{\phi}+\frac{q^{2}}{2}\right)Gdt+\left(-\frac{b}{\phi}+q\right)GdW_{t}.\nonumber
\end{align}
In order to eliminate the noise term we choose $b=q\phi$, so that
$dG=\left(-\phi^{-1}a+p+q^{2}/2\right)Gdt$. We set $q=i(2\kappa)^{1/2}$
and $p=\kappa$, in which case the SDE for $x$ integrates to give
$x(t)=\kappa t+i(2\kappa)^{1/2}W_{t}$, having chosen initial condition
$x(0)=0$. Furthermore, if $a=-\kappa\phi^{2}$ then
\begin{equation}
d\phi=-\kappa\phi^{2}dt+i(2\kappa)^{1/2}\phi dW_{t},\label{eq:116}
\end{equation}
as desired, and $dG=\kappa\phi Gdt=\kappa e^{x}dt=\kappa\exp[\kappa t+i(2\kappa)^{1/2}W_{t}]dt$
which integrates to
\begin{equation}
G(t)=G(0)+\kappa\int_{0}^{t}\exp\left(\kappa s+i(2\kappa)^{1/2}W_{s}\right)ds.\label{eq:118}
\end{equation}
Since $G(t)=\phi^{-1}\exp\left(\kappa t+i(2\kappa)^{1/2}W_{t}\right)$
and $G(0)=\phi_{0}^{-1}$, we then recover Eq. (\ref{eq:119m}).

\section{Formal power series form of noise averaged pseudo-population\label{sec:Formal-power-series}}

We provide the details of the derivation of the formal power series
representation of $\langle\phi(t)\rangle_{\phi_{0}}$ starting from
Eq. (\ref{eq:119m}). The first order term, proportional to $\phi_{0}$,
is straightforward: we write
\begin{equation}
\langle\phi\rangle_{\phi_{0}}\approx\int dW_{t}\,\phi_{0}\,\mathrm{exp}(\kappa t+i(2\kappa)^{\frac{1}{2}}W{}_{t}),\label{eq:126}
\end{equation}
and use the stochastic integral identity $\int dW_{t}\exp(icW_{t})=\langle\exp(icW_{t})\rangle=\exp(-c^{2}t/2)$
such that
\begin{equation}
\langle\phi\rangle_{\phi_{0}}=\phi_{0}+O(\phi_{0}^{2}),\label{eq:127}
\end{equation}
but higher order terms in the formal series
\begin{eqnarray}
\langle\phi\rangle_{\phi_{0}} & = & \sum_{j=1}^{\infty}C_{j}(t)\phi_{0}^{j}=\phi_{0}\sum_{n=0}^{\infty}(-1)^{n}(\kappa\phi_{0})^{n}M_{n}(t)\label{eq:128-1}
\end{eqnarray}
with $C_{j}(t)=(-\kappa)^{j-1}M_{j-1}(t)$ and
\begin{equation}
M_{j}(t)=\left\langle e^{\kappa t+i(2\kappa)^{1/2}W_{t}}\left(\int_{0}^{t}e^{\kappa s+i(2\kappa)^{1/2}W_{s}}ds\right)^{j}\right\rangle .\label{eq:130a}
\end{equation}
require more work.

We focus our attention on the evolution of the mean particle population
$\overline{N}$ for cases where $N_{0}$ particles are present at
$t=0$. We use $P(N,0)=\delta_{NN_{0}}$ with Eqs. (\ref{eq:125})
and (\ref{eq:109}) to obtain
\begin{equation}
\overline{N}=\langle\phi\rangle=\oint_{C}d\phi_{0}\frac{N_{0}!}{2\pi i}\phi_{0}^{-N_{0}-1}\exp(\phi_{0})\langle\phi\rangle_{\phi_{0}},\label{eq:131}
\end{equation}
and by inserting Eq. (\ref{eq:128-1}) and expanding the integrand
we find that
\begin{equation}
\overline{N}=\sum_{j=1}^{N_{0}}C_{j}(t)\frac{N_{0}!}{(N_{0}-j)!}.\label{eq:132}
\end{equation}
Notice that the number of coefficients $C_{j}$ required to evaluate
$\overline{N}$ is finite even though the series for $\langle\phi\rangle_{\phi_{0}}$
is infinite.

We have already established that $C_{1}(t)=1$ in Eq. (\ref{eq:127}).
In order to evaluate $C_{2}(t)$ we need to consider
\begin{equation}
I(t,s)=\left\langle \exp\left(i(2\kappa)^{1/2}\left(W_{t}+W_{s}\right)\right)\right\rangle ,\label{eq:134}
\end{equation}
which in more explicit form reads
\begin{align}
 & I(t,s)=\int_{-\infty}^{\infty}dW_{t}dW_{s}\exp\left(i(2\kappa)^{1/2}\left(W_{t}+W_{s}\right)\right)\nonumber \\
 & \times\frac{1}{(2\pi(t-s))^{1/2}}\exp\left(-\frac{(W_{t}-W_{s})^{2}}{2(t-s)}\right)\label{eq:135}\\
 & \times\frac{1}{(2\pi s)^{1/2}}\exp\left(-\frac{W_{s}^{2}}{2s}\right),\nonumber
\end{align}
in which the weighting is a product of the gaussian probabilities
for generating a value $W_{s}$ of the Wiener process at time $s$,
and a value $W_{t}$ at the later time $t$ given the earlier value.
Writing $W_{t}+W_{s}=W_{t}^{\prime}+2W_{s}$ where $W_{t}^{\prime}=W_{t}-W_{s}$
this factorises as follows:
\begin{eqnarray}
I(t,s) & = & \int dW_{t}^{\prime}\exp\left(i(2\kappa)^{1/2}W_{t}^{\prime}\right)\nonumber \\
 & \times & \frac{1}{(2\pi(t-s))^{1/2}}\exp\left(-\frac{W_{t}^{\prime2}}{2(t-s)}\right)\nonumber \\
 & \times & \int dW_{s}\exp(2i(2\kappa)^{1/2}W_{s})\frac{1}{(2\pi s)^{1/2}}\exp\left(-\frac{W_{s}^{2}}{2s}\right)\nonumber \\
 & = & \langle\exp(i(2\kappa)^{1/2}W_{t-s})\rangle\langle\exp(2i(2\kappa)^{1/2}W_{s})\rangle\nonumber \\
 & = & e^{-\kappa(t-s)}e^{-4\kappa s}=e^{-\kappa(t+3s)},\label{eq:136}
\end{eqnarray}
using $\langle\exp(icW_{t})\rangle=\exp(-c^{2}t/2)$, and hence $C_{2}(t)=-\kappa M_{1}(t)$
with
\begin{equation}
\begin{aligned} & M_{1}(t)=e^{\kappa t}\int_{0}^{t}ds\, e^{\kappa s}\left\langle \exp\left(i(2\kappa)^{1/2}\left(W_{t}+W_{s}\right)\right)\right\rangle \\
 & =e^{\kappa t}\int_{0}^{t}ds\, e^{\kappa s}I(t,s)=\int_{0}^{t}ds\, e^{-2\kappa s}=\frac{1}{2\kappa}\left(1-e^{-2\kappa t}\right).
\end{aligned}
\label{eq:137}
\end{equation}
 Consider next $C_{3}(t)=\kappa^{2}M_{2}(t)$. We need to evaluate
\begin{equation}
\begin{aligned} & M_{2}=\int_{0}^{t}ds_{1}\int_{0}^{t}ds_{2}e^{\kappa(t+s_{1}+s_{2})}\\
 & \times\left\langle \exp\left[i(2\kappa)^{1/2}\left(W_{t}+W_{s_{1}}+W_{s_{2}}\right)\right]\right\rangle \\
 & =2\int_{0}^{t}ds_{1}\int_{0}^{s_{1}}ds_{2}e^{\kappa(t+s_{1}+s_{2})}\\
 & \times\left\langle \exp\left[i(2\kappa)^{1/2}\left(W_{t}-W_{s_{1}}+2(W_{s_{1}}-W_{s_{2}})+3W_{s_{2}}\right)\right]\right\rangle ,
\end{aligned}
\label{eq:138}
\end{equation}
where an ordering $t\ge s_{1}\ge s_{2}\ge0$ has been imposed by the
choice of integration limits, the change in which is accounted for
by inserting the prefactor of two. Once again this reduces to
\begin{equation}
\begin{aligned} & M_{2}=2\int_{0}^{t}ds_{1}\int_{0}^{s_{1}}ds_{2}\exp(\kappa(t+s_{1}+s_{2}))\\
 & \times\exp\left(-\kappa(t-s_{1})\right)\exp\left(-4\kappa(s_{1}-s_{2})\right)\exp\left(-9\kappa s_{2}\right)\\
 & =2\int_{0}^{t}ds_{1}\int_{0}^{s_{1}}ds_{2}\exp(-2\kappa s_{1})\exp(-4\kappa s_{2}),
\end{aligned}
\label{eq:139}
\end{equation}
which becomes
\begin{eqnarray}
M_{2} & = & \frac{1}{2\kappa}\int_{0}^{t}ds_{1}\exp(-2\kappa s_{1})\left(1-\exp(-4\kappa s_{1})\right)\nonumber \\
 & = & \frac{1}{2\kappa}\left[\frac{1}{2\kappa}\left(1-e^{-2\kappa t}\right)-\frac{1}{6\kappa}\left(1-e^{-6\kappa t}\right)\right].\label{eq:140}
\end{eqnarray}
Similarly, we can show that
\begin{equation}
M_{3}=6\int_{0}^{t}ds_{1}\int_{0}^{s_{1}}ds_{2}\int_{0}^{s_{2}}ds_{3}e^{-2\kappa s_{1}}e^{-4\kappa s_{2}}e^{-6\kappa s_{3}},\label{eq:141}
\end{equation}
and the pattern that emerges is
\begin{equation}
M_{n}=n!\int_{0}^{t}ds_{1}\cdots\int_{0}^{s_{n-1}}ds_{n}e^{-2\kappa s_{1}}\cdots e^{-2n\kappa s_{n}}.\label{eq:142}
\end{equation}
We notice that the $M_{n}$ are related to one another by repeated
integration of the nested integrals. We define
\begin{equation}
\begin{aligned} & m_{n}(\kappa t)=\frac{(2\kappa)^{n}}{n!}M_{n}(t)\\
 & =\int_{0}^{2\kappa t}ds_{1}\int_{0}^{s_{1}}ds_{2}\cdots\int_{0}^{s_{n-1}}ds_{n}e^{-s_{1}}e^{-2s_{2}}\cdots e^{-ns_{n}},
\end{aligned}
\label{eq:143a}
\end{equation}
such that $m_{1}=1-\exp(-2\kappa t)$ and
\begin{equation}
m_{2}=\frac{1}{6}\left(2-3e^{-2\kappa t}+e^{-6\kappa t}\right)=\frac{1}{2}m_{1}-\frac{1}{6}\left(1-e^{-6\kappa t}\right).\label{eq:144}
\end{equation}
Similarly we can show that
\begin{equation}
m_{3}=\frac{1}{3}m_{2}-\frac{1}{3}\frac{1}{5}m_{1}+\frac{1}{3}\frac{1}{5}\frac{1}{6}\left(1-e^{-12\kappa t}\right).\label{eq:145a}
\end{equation}
On the basis of an analysis of further cases, we conjecture a pattern
\begin{equation}
\begin{aligned} & m_{n}=\frac{1}{n}m_{n-1}-\frac{m_{n-2}}{n[n+(n-1)]}\\
 & +\frac{m_{n-3}}{n[n+(n-1)][n+(n-1)+(n-2)]}\\
 & \cdots+(-1)^{n}\frac{m_{1}}{n\left[n+(n-1)\right]\cdots\left[n+(n-1)+\cdots+2\right]}\\
 & \!\!+(-1)^{n+1}\frac{1}{n\left[n+(n-1)\right]\cdots\left[n+(n-1)+\cdots+2\right]}\\
 & \!\!\times\!\int_{0}^{2\kappa t}\!\!\! ds_{1}\exp[-(n+[n-1]+\cdots+1)s_{1}].
\end{aligned}
\label{eq:146}
\end{equation}
This may be simplified since $n+(n-1)+\cdots+1=\frac{1}{2}n(n+1)$
is a triangular number, and
\begin{equation}
\begin{aligned} & n+(n-1)+\cdots+(n-j+1)\\
 & =n+(n-1)+\cdots+1\\
 & -[(n-j)+(n-j-1)+\cdots+1]\\
 & =\frac{1}{2}n(n+1)-\frac{1}{2}(n-j)(n-j+1)=\frac{j}{2}\left(2n+1-j\right),
\end{aligned}
\label{eq:147}
\end{equation}
is a difference of triangular numbers, such that
\begin{eqnarray}
m_{n} & = & \frac{1}{n}m_{n-1}-\frac{1}{n[n+(n-1)]}m_{n-2}+\cdots\nonumber \\
 &  & +(-1)^{l+1}\left(\prod_{j=1}^{l}\frac{j}{2}\left(2n+1-j\right)\right)^{-1}m_{n-l}+\cdots\nonumber \\
 &  & +(-1)^{n+1}\left(\prod_{j=1}^{n}\frac{j}{2}\left(2n+1-j\right)\right)^{-1}m_{0}(n),\label{eq:148}
\end{eqnarray}
where we define
\begin{equation}
m_{0}(n)=1-\exp[-n(n+1)\kappa t].\label{eq:149-1}
\end{equation}
Since
\begin{equation}
\begin{aligned} & \prod_{j=1}^{l}\frac{j}{2}\left(2n+1-j\right)\\
 & =\frac{1}{2^{l}}\left(1.2.\cdots l\right)\left[2n\cdots(2n+1-l)\right]=\frac{l!(2n)!}{2^{l}(2n-l)!},
\end{aligned}
\label{eq:150}
\end{equation}
we finally obtain
\begin{equation}
m_{n}=\sum_{l=1}^{n}\frac{2^{l}(2n-l)!(-1)^{l+1}}{(2n)!l!}m_{n-l},\label{eq:151a}
\end{equation}
in agreement with Eqs. (\ref{eq:144}) and (\ref{eq:145a}), and where
the $n$-dependence of $m_{0}$ appearing in the final term is understood.
Evaluation of the $m_{n}$ by integration of Eq. (\ref{eq:143a})
using \emph{Mathematica} \cite{Mathematica} for $n$ up to twenty
confirms this relation and the conjectured pattern that relates them.

\section{Mean and variance of population in example cases\label{sec:Mean-populations-in}}

We found in Appendix \ref{sec:Formal-power-series} that $C_{1}(t)=1$
and
\begin{eqnarray}
C_{2}(t) & = & -\frac{1}{2}m_{1}=-\frac{1}{2}\left(1-e^{-2\kappa t}\right)\nonumber \\
C_{3}(t) & = & \frac{1}{2}m_{2}=\frac{1}{12}\left(2-3e^{-2\kappa t}+e^{-6\kappa t}\right)\label{eq:152}\\
C_{4}(t) & = & \frac{3}{4}m_{3}=-\frac{1}{24}+\frac{3}{40}e^{-2\kappa t}-\frac{1}{24}e^{-6\kappa t}+\frac{1}{120}e^{-12\kappa t},\nonumber
\end{eqnarray}
and we use these and Eq. (\ref{eq:132}) to obtain the exact time
dependence of the mean population for values of initial population
$N_{0}$ from 1 to 4. For $N_{0}=1$, we obtain
\begin{equation}
\overline{N}=C_{1}(t)\frac{N_{0}!}{(N_{0}-1)!}=1,\label{eq:153}
\end{equation}
as expected. For $N_{0}=2$ we find that
\begin{eqnarray}
\overline{N} & = & \sum_{j=1}^{2}C_{j}(t)\frac{N_{0}!}{(N_{0}-j)!}=2C_{1}(t)+2C_{2}(t)\nonumber \\
 & = & 2-\left(1-e^{-2\kappa t}\right)=1+e^{-2\kappa t},\label{eq:154}
\end{eqnarray}
and this can be checked by solving the underlying master equations.
 For $N_{0}=3$ we get
\begin{eqnarray}
\overline{N} & = & \sum_{j=1}^{3}C_{j}(t)\frac{N_{0}!}{(N_{0}-j)!}=3C_{1}(t)+6C_{2}(t)+6C_{3}(t)\nonumber \\
 & = & 1+\frac{3}{2}e^{-2\kappa t}+\frac{1}{2}e^{-6\kappa t},\label{eq:155}
\end{eqnarray}
which may also be checked by solving the master equations directly.
Finally for $N_{0}=4$ we obtain
\begin{eqnarray}
\overline{N} & = & \sum_{j=1}^{4}C_{j}(t)\frac{N_{0}!}{(N_{0}-j)!}\nonumber \\
 & = & 4C_{1}(t)+12C_{2}(t)+24C_{3}(t)+24C_{4}(t)\nonumber \\
 & = & 1+\frac{9}{5}e^{-2\kappa t}+e^{-6\kappa t}+\frac{1}{5}e^{-12\kappa t}.\label{eq:156}
\end{eqnarray}
All these solutions satisfy the initial condition $\overline{N}=N_{0}$
at $t=0$ and tend to unity as $t\rightarrow\infty$, as required.

These results may be compared with those of Barzykin and Tachiya \cite{Barzykin05}
who obtained an exact solution to the master equations for this problem
using a generating function approach. They found that
\begin{equation}
\overline{N}=\sum_{j=1}^{N_{0}}(2j-1)\frac{(N_{0}-1)!}{(N_{0}+j-1)!}\frac{N_{0}!}{(N_{0}-j)!}e^{-j(j-1)\kappa t)},\label{eq:157}
\end{equation}
and our calculations are consistent with this expression.

We can also determine the variance in population as a function of
time and initial condition $N_{0}$. We need the second moment $\overline{N^{2}}$,
which is equivalent to the quantity $\langle\phi^{2}+\phi\rangle$.
We could expand $\phi^{2}$ as a formal power series, just as we did
for $\phi$ in Appendix \ref{sec:Formal-power-series}, but instead
we exploit the relationship $d\langle\phi\rangle/dt=-\kappa\langle\phi^{2}\rangle$
that arises from Eq. (\ref{eq:112}). We hence obtain from Eq. (\ref{eq:132}):
\begin{equation}
\overline{N^{2}}=\sum_{j=1}^{N_{0}}\left(-\frac{1}{\kappa}\frac{dC_{j}(t)}{dt}+C_{j}(t)\right)\frac{N_{0}!}{(N_{0}-j)!},\label{eq:160}
\end{equation}
and using Eqs. (\ref{eq:152}) we can construct the time derivatives,
namely $dC_{1}(t)/dt=0$ together with
\begin{eqnarray}
\frac{1}{\kappa}\frac{dC_{2}(t)}{dt} & = & -e^{-2\kappa t}\nonumber \\
\frac{1}{\kappa}\frac{dC_{3}(t)}{dt} & = & \frac{1}{2}\left(e^{-2\kappa t}-e^{-6\kappa t}\right)\label{eq:161}\\
\frac{1}{\kappa}\frac{dC_{4}(t)}{dt} & = & -\frac{3}{20}e^{-2\kappa t}+\frac{1}{4}e^{-6\kappa t}-\frac{1}{10}e^{-12\kappa t},\nonumber
\end{eqnarray}
which puts us in a position to calculate the variance $\sigma^{2}=\overline{N^{2}}-\overline{N}^{2}$.

Considering first the trivial case $N_{0}=1$, we deduce $\overline{N^{2}}=1$
and $\sigma^{2}=0$ as we would expect. For $N_{0}=2$, $\overline{N^{2}}=2+2\left[\exp\left(-2\kappa t\right)-\frac{1}{2}\left(1-\exp\left(-2\kappa t\right)\right)\right]=1+3\exp\left(-2\kappa t\right)$.
This has the correct limits of $\overline{N^{2}}=4$ and 1 for $t=0$
and $t\rightarrow\infty$, respectively. The variance is $\sigma^{2}=1+3\exp\left(-2\kappa t\right)-\left(1+\exp\left(-2\kappa t\right)\right)^{2}=\exp\left(-2\kappa t\right)-\exp\left(-4\kappa t\right)$
. This has the correct limits of zero at both $t=0$ and $t\rightarrow\infty$
and reaches a maximum value of $1/4$ when $-2\kappa+4\kappa\exp(-2\kappa t)=0$
or $t=(2\kappa)^{-1}\ln2$.

\begin{figure}
\begin{centering}
\includegraphics[width=1\columnwidth]{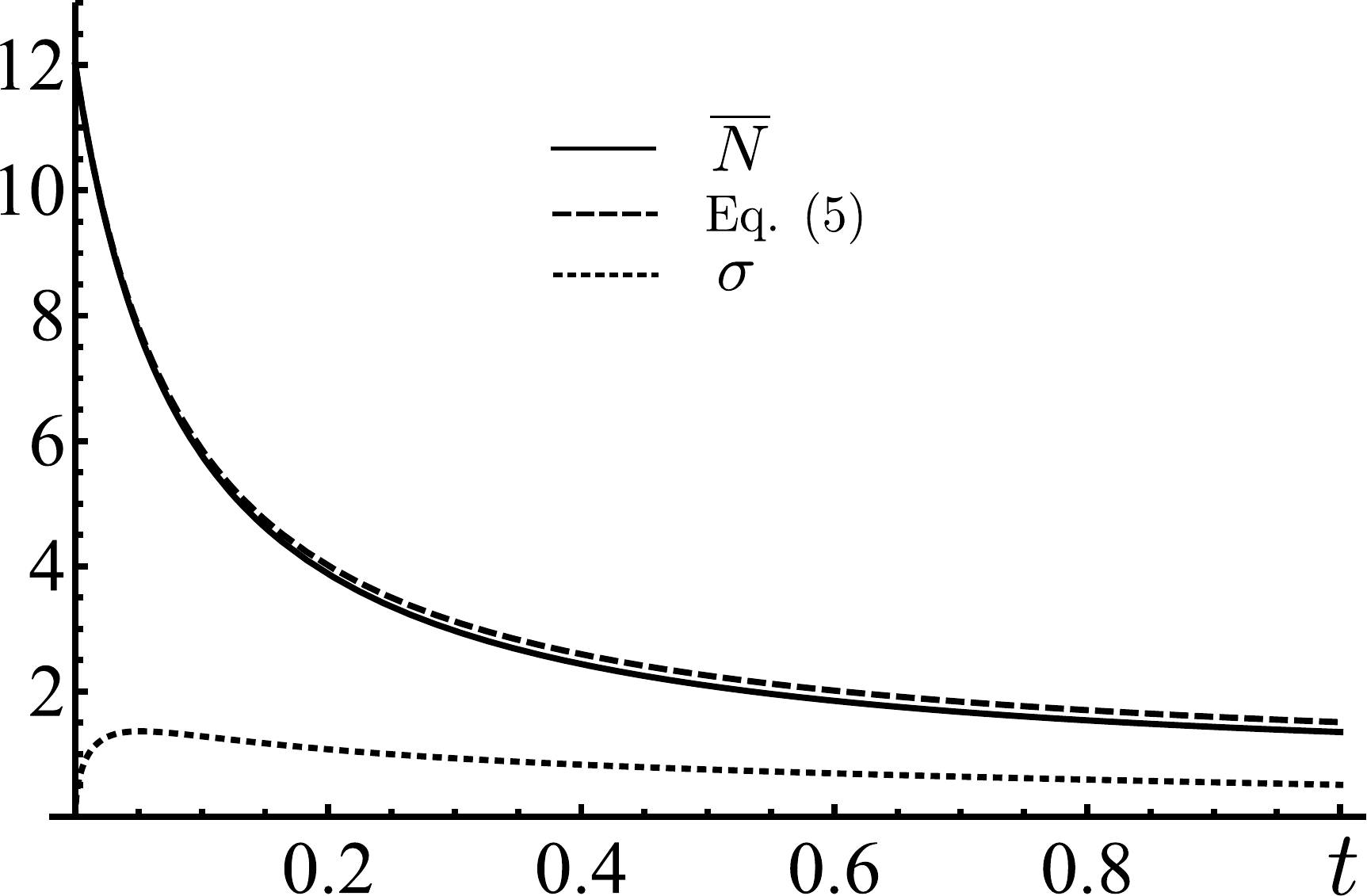}
\par\end{centering}

\protect\caption{Mean (solid line) and standard deviation (short dashed line) of the
population as a function of time, in a coagulating system with initial
population $N_{0}=12$ and with $\kappa=1$. The long dashed curve
is the mean population according to Eq. (\ref{eq:103}). \label{fig:Mean-(red)-and}}
\end{figure}

We have employed \emph{Mathematica} \cite{Mathematica} to calculate
higher order coefficients $m_{n}$ and hence the $C_{n}(t)$ in Eq.
(\ref{eq:132}). We illustrate the outcome by computing the time-dependent
mean and standard deviation for the case with $N_{0}=12$, shown in
Figure \ref{fig:Mean-(red)-and}. The exact form of $\overline{N}$
obtained from the analysis is
\begin{equation}
\begin{aligned} & \overline{N}=\frac{1}{58786}\biggl(e^{-132\kappa t}\!+21e^{-110\kappa t}\!+209e^{-90\kappa t}\!+1309e^{-72\kappa t}\\
 & \!+5775e^{-56\kappa t}\!+19019e^{-42\kappa t}\!+48279e^{-30\kappa t}\!+95931e^{-20\kappa t}\\
 & \!+149226e^{-12\kappa t}+177650e^{-6\kappa t}+149226e^{-2\kappa t}+58786\biggr),
\end{aligned}
\label{eq:162}
\end{equation}
which is consistent with the Barzykin and Tachiya expression \cite{Barzykin05}.
Eq. (\ref{eq:103}) happens to account reasonably well for the evolution
of the mean population, as shown in Figure \ref{fig:Mean-(red)-and},
but of course the mean field approximation upon which it is based
cannot reproduce the standard deviation.

\section{Steepest descent evaluation of $\overline{N}$ \label{sec:Steepest-descent-evaluation}}

If we employ the representation Eq. (\ref{eq:125}) the set of initial
values of $\phi_{0}$ for the stochastic dynamics should form a closed
contour around the origin in the complex plane. We now demonstrate
that for the purposes of computation there is an optimal contour,
depending on the initial population in the problem $N_{0}$, and that
\begin{equation}
\overline{N}=\oint_{C}d\phi_{0}f_{N_{0}}(\phi_{0})\langle\phi\rangle_{\phi_{0}},\label{eq:a1}
\end{equation}
where $f_{N_{0}}(\phi_{0})$ is given by Eq. (\ref{eq:109}), reduces
to Eq. (\ref{eq:102}) for situations with a definite initial population,
at least for small $\kappa t$.

Contour integrals of the form $\oint\exp[q(z)]dz$ may be evaluated
by the method of steepest descents about saddle points in $q(z)$
defined by locations $z=z_{s}$ in the complex plane where $q^{\prime}=dq(z)/dz=0$.
We represent the integrand near a saddle point as $\exp(q(z))\approx\exp(q(z_{s}))\exp(q^{\prime\prime}(z_{s})(z-z_{s})^{2}/2)$
and integrate along a contour through $z_{s}$ on which the imaginary
part of $q^{\prime\prime}(z_{s})(z-z_{s})^{2}$ is zero and the real
part is negative. If the contour is arranged such that it passes over
saddle points along the line of steepest descent on the surface of
the modulus of the integrand, and elsewhere follows valleys where
the modulus of the integrand is small, then the contour integral may
be approximated by a set of gaussian integrations about each saddle
point position.

\begin{figure}
\includegraphics[width=1\columnwidth]{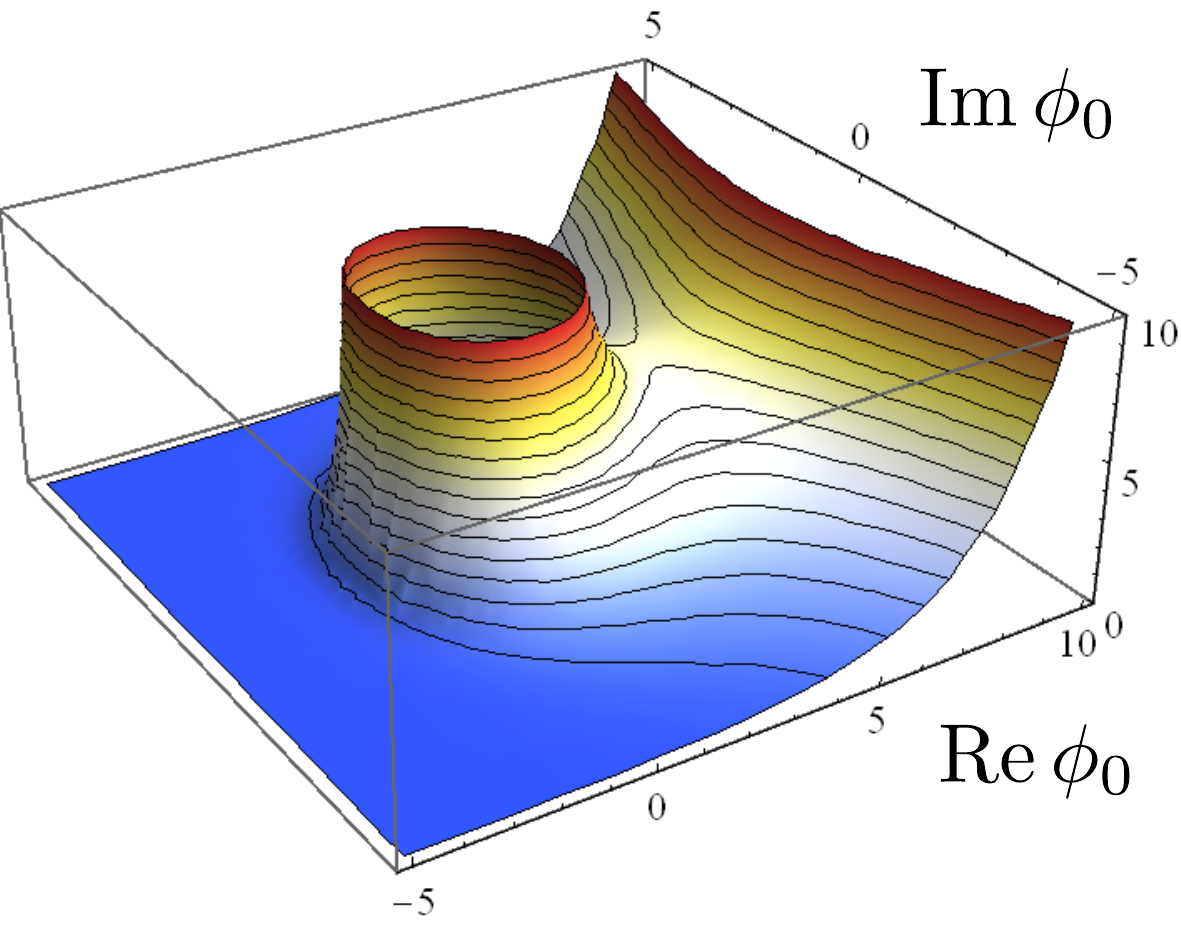}

\protect\caption{Modulus of $\phi_{0}^{-N_{0}}\exp(\phi_{0})$ for $N_{0}=5$, indicating
the saddle point at $\phi_{0}=5$: the path of steepest descent through
this point lies parallel to the imaginary axis.\label{fig:Modulus-of-}}
\end{figure}

We write
\begin{equation}
\overline{N}=\oint_{C}d\phi_{0}\frac{N_{0}!}{2\pi i}\exp\left(\phi_{0}-(N_{0}+1)\ln\phi_{0}+\ln\langle\phi\rangle_{\phi_{0}}\right),\label{eq:a2}
\end{equation}
and consider $\kappa t\ll1$. In this limit $M_{n}\approx t^{n}$,
according to Eq. (\ref{eq:142}), and hence $C_{j}(t)\approx(-\kappa t)^{j-1}$
such that $\langle\phi\rangle_{\phi_{0}}\approx\phi_{0}+O(\kappa t)$.
We shall write $\langle\phi\rangle_{\phi_{0}}=\phi_{0}F(\phi_{0})$
with $F$ only weakly dependent on $\phi_{0}$, and seek saddle points
in $q(\phi_{0})=\phi_{0}-N_{0}\ln\phi_{0}$. Solving $q^{\prime}(\phi_{0})=1-N_{0}/\phi_{0}=0$
identifies a single saddle point of the integrand at $\phi_{0}=N_{0}$.
The path of steepest descent passes through the saddle point perpendicular
to the real axis. The modulus of the integrand, for $N_{0}=5$, is
shown in Figure \ref{fig:Modulus-of-}.

Clearly, it is sensible to place the contour $C$ along the path of
steepest descent through the saddle point and then to complete the
circuit around the origin through regions where the modulus of the
integrand is small. The contour integral is approximated by the contribution
along a line parallel to the imaginary axis through $\phi_{0}=N_{0}.$
Writing $\phi_{0}=N_{0}+iy$ we have

\begin{equation}
\begin{aligned} & \overline{N}\approx\exp\left(\ln F(N_{0})\right)\\
 & \times\int_{-\infty}^{\infty}i\, dy\frac{N_{0}!}{2\pi i}\exp\left(q(N_{0})+\frac{1}{2}q^{\prime\prime}(N_{0})(\phi_{0}-N_{0})^{2}\right)\\
 & \approx\frac{N_{0}!}{2\pi}\langle\phi\rangle_{N_{0}}N_{0}^{-1}\int_{-\infty}^{\infty}dy\: e^{N_{0}-N_{0}\ln N_{0}+(iy)^{2}/(2N_{0})}\\
 & \approx\langle\phi\rangle_{N_{0}}N_{0}!\left(\frac{N_{0}}{2\pi}\right)^{1/2}\frac{\exp(N_{0})}{N_{0}^{N_{0}+1}}\approx\langle\phi\rangle_{N_{0}},
\end{aligned}
\label{eq:a3}
\end{equation}
using $n!\approx n^{n}\exp(-n)(2\pi n)^{1/2}$ which is quite accurate
even for $N_{0}$ as low as unity. Finally, the expansion (\ref{eq:128})
with the $C_{j}(t)$ valid for $\kappa t\ll1$ may be written
\begin{equation}
\langle\phi\rangle_{\phi_{0}}\approx\sum_{j=1}^{\infty}(-\kappa t)^{j-1}\phi_{0}^{j}=\frac{\phi_{0}}{1+\kappa\phi_{0}t},\label{eq:a4}
\end{equation}
making Eq. (\ref{eq:a3}) and hence Eq. (\ref{eq:a2}) consistent
with the mean field approximation $\overline{N}\approx N_{0}/(1+\kappa N_{0}t)$
for small $\kappa t$.

\section{Equivalence of Drummond gauging and the Cameron-Martin-Girsanov formula\label{sec:Equivalence-of-Drummond}}

The equivalence between the statistical properties of the gauged stochastic
variable $\phi^{\prime}$, when suitably weighted, and those of the
ungauged variable $\phi$, derived by Drummond \cite{Drummond04}
and explored in Section \ref{sub:Elimination-of-instabilities}, is
a consequence of some fundamental rules in stochastic calculus that
are expressed by the Cameron-Martin-Girsanov formula \cite{CameronMartin44,CameronMartin45,Girsanov60,BaxterRennie96}.
An SDE such as
\begin{equation}
dx=adt+bdW_{t},\label{eq:301}
\end{equation}
is a statement of a connection between $dx$ and a stochastic variable
$dW_{t}$ with certain statistical properties. When we evaluate expectation
values such as $\langle dx\rangle$ we are implicitly defining a probability
distribution or\emph{ }measure\emph{ }over the values taken by the
variable $dW_{t}$. Normally the notation $dW_{t}$ represents an
increment in a Wiener process and the implication is that the probability
distribution of values of $dW_{t}$ is gaussian with zero mean and
variance equal to $dt$.

But how might the expectation value of the increment $dW_{t}$ change
if we were to evaluate it with respect to a different probability
distribution? For example, what if it were distributed according to
a shifted gaussian proportional to $\exp\left(-(dW_{t}-m)^{2}/\left(2dt\right)\right)$
where $m$ is the non-zero mean of $dW_{t}$ making it no longer an
increment in a Wiener process? Let us take the mean of $dW_{t}$ under
such a shifted gaussian distribution to be proportional to the time
elapsed, such that we write $m=\mu dt$ and $\langle dW_{t}\rangle_{Q}=\mu dt$.
The new probability distribution, denoted $Q$, is indicated through
a suffix on the expectation value. We write $\langle dW_{t}\rangle_{P}=0$
in the old measure which we denote $P$, according to which $dW_{t}$
is indeed an increment in a Wiener process.

Note that $\langle dW_{t}-\mu dt\rangle_{Q}=0$ and $\langle\left(dW_{t}-\mu dt\right)^{2}\rangle_{Q}=dt$
since the variance of $dW_{t}$ under measure $Q$ is the same as
that under $P$: we have shifted the gaussian probability distribution
for $dW_{t}$ but not changed its width. Hence, if we define $d\tilde{W_{t}}=dW_{t}-\mu dt$
then $\langle d\tilde{W}_{t}\rangle_{Q}=0$ and $\langle(d\tilde{W}_{t})^{2}\rangle_{Q}=dt$:
we can therefore identify a Wiener process that operates under probability
measure $Q$, and relate it to a Wiener process under measure $P$.
The SDE for $x$ now reads
\begin{equation}
dx=(a+b\mu)dt+bd\tilde{W}_{t},\label{eq:302}
\end{equation}
 and we can choose to evaluate expectation values under measure $P,$
for which $\langle d\tilde{W}_{t}\rangle_{P}=-\mu dt$, or measure
$Q,$ for which $\langle d\tilde{W}_{t}\rangle_{Q}=0$.

The expectation value $\langle dx\rangle_{Q}$ is the solution to
a problem that differs from the one initially posed, since the drift
term in the SDE has been changed from $adt$ to $(a+b\mu)dt$. However,
the point is that it is possible to establish a link between expectation
values under the two different measures. The average of $dx$ under
measure $P$ is equal to a \emph{weighted} average of $dx$ under
$Q$. Formally, we can write
\begin{equation}
\langle dx\rangle_{P}=\left\langle \frac{dP}{dQ}dx\right\rangle _{Q},\label{eq:303}
\end{equation}
where $dP/dQ$ is the Radon-Nikodym derivative of probability measure
$P$ with respect to $Q$. Furthermore, the Cameron-Martin-Girsanov
formula states that we can write
\begin{equation}
\frac{dP}{dQ}=\exp\left(-\mu d\tilde{W}_{t}-\frac{1}{2}\mu^{2}dt\right),\label{eq:304}
\end{equation}
which is just a ratio of the gaussian distributions $\exp[-(d\tilde{W}_{t}+\mu dt)^{2}/\left(2dt\right)]$
and $\exp[-d\tilde{W}_{t}{}^{2}/\left(2dt\right)]$. We now define
a quantity $\Omega$ through the relation $\exp(d\ln\Omega)=dP/dQ$
such that
\begin{equation}
d\ln\Omega=-\mu^{2}dt/2-\mu d\tilde{W}_{t},\label{eq:306}
\end{equation}
 and using Ito's lemma, it can be shown that $\Omega$ evolves according
to
\begin{equation}
\begin{aligned} & d\Omega=d\exp\left(\ln\Omega\right)=\exp\left(\ln\Omega\right)d\ln\Omega+\frac{1}{2}\exp\left(\ln\Omega\right)\mu^{2}dt\\
 & =\Omega\left(-\mu^{2}dt/2-\mu d\tilde{W}_{t}+\mu^{2}dt/2\right)=-\mu\Omega d\tilde{W}_{t}.
\end{aligned}
\label{eq:307}
\end{equation}
We see from Eqs. (\ref{eq:303}-\ref{eq:307}) that averages of the
increment $dx$ over differently distributed random increments can
be related to one another, and that the quantity $d\Omega$ describes
the connection. In order for this to make sense mathematically, two
conditions must be met. The first is that no process that has non-zero
probability under $P$ should be impossible under $Q$, and vice versa.
The second, known as Novikov's condition, requires that $\left\langle \exp\left(\frac{1}{2}\mu^{2}dt\right)\right\rangle <\infty$.

We now invert the point of view, and instead of considering an increment
of a variable treated according to two different averaging procedures,
we relate increments of two different variables under the same averaging.
Explicitly, we consider variable $x$ evolving as $dx=adt+bdW_{t}$
and another variable $x^{\prime}$ that evolves according to $dx^{\prime}=(a+b\mu)dt+bdW_{t}$,
with $x(0)=x^{\prime}(0)$. The above results imply that we can write
\begin{equation}
\langle dx\rangle=\langle\exp(-\mu^{2}dt/2-\mu dW_{t})dx^{\prime}\rangle,\label{eq:308}
\end{equation}
which resembles a combination of Eqs. (\ref{eq:303}) and (\ref{eq:304}).
We shall demonstrate its validity shortly. Note that there is an implied
suffix $P$ on the brackets, since $dW_{t}$ is here a standard Wiener
increment with zero mean. The factor inserted in front of $dx^{\prime}$
accounts for the difference in drift in the evolution of $x$ and
$x^{\prime}$ and may also be written $\exp(d\ln\Omega)$ with $d\ln\Omega=-\mu^{2}dt/2-\mu dW_{t}$
or equivalently $d\Omega=-\mu\Omega dW_{t}$. Clearly we can write
$\exp(\int d\ln\Omega)=\exp[\ln\Omega(t)-\ln\Omega(0)]=\Omega(t)/\Omega(0)$
and if we set $\Omega(0)=1$ then
\begin{equation}
\begin{aligned}\Omega(t)=\exp\left(-\frac{1}{2}\int_{0}^{t}\mu^{2}(t^{\prime})dt^{\prime}-\int_{0}^{t}\mu(t^{\prime})dW_{t^{\prime}}\right).\end{aligned}
\label{eq:309}
\end{equation}
Note that $\langle\Omega\rangle=\Omega(0)+\int\langle d\Omega\rangle=1-\int\langle\mu\Omega dW_{t}\rangle=1$.
And for a finite interval of time, Novikov's condition on $\mu$ reads
$\left\langle \exp\left(\frac{1}{2}\int_{0}^{t}\mu^{2}(t^{\prime})dt^{\prime}\right)\right\rangle <\infty$.

Now consider the following:
\begin{equation}
\begin{aligned} & \left\langle \Omega(t)\left(x^{\prime}(t)-x^{\prime}(0)\right)\right\rangle \\
 & =\left\langle \prod_{i}e^{-\frac{1}{2}\mu^{2}(t_{i})dt-\mu(t_{i})dW_{i}}\sum_{j}dx_{j}^{\prime}\right\rangle \\
 & =\sum_{j}\left\langle e^{-\frac{1}{2}\mu^{2}(t_{j})dt-\mu(t_{j})dW_{j}}dx_{j}^{\prime}\right\rangle ,
\end{aligned}
\label{eq:310}
\end{equation}
 where we use a discrete representation of the time integration. In
deriving this we have recognised that if $k\ne j$ then
\begin{equation}
\begin{aligned} & \!\!\!\left\langle \!\! e^{-\frac{1}{2}\mu^{2}(t_{k})dt-\mu(t_{k})dW_{k}}\prod_{i\ne k}e^{-\frac{1}{2}\mu^{2}(t_{i})dt-\mu(t_{i})dW_{i}}dx_{j}^{\prime}\right\rangle \\
 & =\left\langle e^{-\frac{1}{2}\mu^{2}(t_{k})dt-\mu(t_{k})dW_{k}}\right\rangle \left\langle \prod_{i\ne k}e^{-\frac{1}{2}\mu^{2}(t_{i})dt-\mu(t_{i})dW_{i}}dx_{j}^{\prime}\right\rangle \\
 & =\left\langle \prod_{i\ne k}e^{-\frac{1}{2}\mu^{2}(t_{i})dt-\mu(t_{i})dW_{i}}dx_{j}^{\prime}\right\rangle ,
\end{aligned}
\label{eq:311}
\end{equation}
such that by repetition of this step, the right hand side reduces
to $\left\langle \exp[-\frac{1}{2}\mu^{2}(t_{j})dt-\mu(t_{j})dW_{j}]dx_{j}^{\prime}\right\rangle $.

Using $\langle\exp\left(-\mu dW_{t}\right)\rangle=\exp(\mu^{2}dt/2)$
we can write
\begin{eqnarray}
\left\langle \exp\left(-\mu dW_{t}\right)dW_{t}\right\rangle  & = & -d\left\langle \exp\left(-\mu dW_{t}\right)\right\rangle /d\mu\nonumber \\
 & = & -\mu dt\,\exp(\mu^{2}dt/2),\label{eq:311a}
\end{eqnarray}
 and with the insertion of $dx^{\prime}=(a+b\mu)dt+bdW_{t}$, we conclude
that
\begin{equation}
\begin{aligned} & \left\langle e^{-\frac{1}{2}\mu^{2}(t_{j})dt-\mu(t_{j})dW_{j}}dx_{j}^{\prime}\right\rangle =\\
 & \left(a(t_{j})+b\mu(t_{j})\right)dt-b\mu(t_{j})dt=a(t_{j})dt=\langle dx_{j}\rangle,
\end{aligned}
\label{eq:312}
\end{equation}
such that
\begin{equation}
\!\left\langle \Omega(t)\!\left(x^{\prime}(t)-x^{\prime}(0)\right)\right\rangle =\sum_{j}\langle dx_{j}\rangle=\left\langle x(t)-x(0)\right\rangle ,\label{eq:312a}
\end{equation}
and since
\begin{equation}
\left\langle \Omega(t)x^{\prime}(0)\right\rangle =\left\langle \Omega(t)\right\rangle \left\langle x^{\prime}(0)\right\rangle =\left\langle x(0)\right\rangle ,\label{eq:312b}
\end{equation}
this means that
\begin{equation}
\langle x(t)\rangle=\left\langle \Omega(t)x^{\prime}(t)\right\rangle .\label{eq:313}
\end{equation}
Thus we have shown that if we wish to evaluate the quantity $\langle x(t)\rangle$
generated by SDE $dx=adt+bdW_{t}$, we could instead solve the SDEs
\begin{eqnarray}
dx^{\prime} & = & adt+b(dW_{t}-gdt)\nonumber \\
d\Omega & = & g\Omega dW_{t},\label{eq:314}
\end{eqnarray}
with initial conditions $x^{\prime}(0)=x(0)$, $\Omega(0)=1$, and
with the function $g(t)=-\mu(t)$ taking arbitrary form subject to
Novikov's condition, and then use the results to evaluate the equivalent
quantity $\langle\Omega(t)x^{\prime}(t)\rangle$. Clearly, this is
identical to Drummond's gauging scheme.

\bibliography{coag}

\begin{thebibliography}{43}%
\makeatletter
\providecommand \@ifxundefined [1]{%
 \@ifx{#1\undefined}
}%
\providecommand \@ifnum [1]{%
 \ifnum #1\expandafter \@firstoftwo
 \else \expandafter \@secondoftwo
 \fi
}%
\providecommand \@ifx [1]{%
 \ifx #1\expandafter \@firstoftwo
 \else \expandafter \@secondoftwo
 \fi
}%
\providecommand \natexlab [1]{#1}%
\providecommand \enquote  [1]{``#1''}%
\providecommand \bibnamefont  [1]{#1}%
\providecommand \bibfnamefont [1]{#1}%
\providecommand \citenamefont [1]{#1}%
\providecommand \href@noop [0]{\@secondoftwo}%
\providecommand \href [0]{\begingroup \@sanitize@url \@href}%
\providecommand \@href[1]{\@@startlink{#1}\@@href}%
\providecommand \@@href[1]{\endgroup#1\@@endlink}%
\providecommand \@sanitize@url [0]{\catcode `\\12\catcode `\$12\catcode
  `\&12\catcode `\#12\catcode `\^12\catcode `\_12\catcode `\%12\relax}%
\providecommand \@@startlink[1]{}%
\providecommand \@@endlink[0]{}%
\providecommand \url  [0]{\begingroup\@sanitize@url \@url }%
\providecommand \@url [1]{\endgroup\@href {#1}{\urlprefix }}%
\providecommand \urlprefix  [0]{URL }%
\providecommand \Eprint [0]{\href }%
\providecommand \doibase [0]{http://dx.doi.org/}%
\providecommand \selectlanguage [0]{\@gobble}%
\providecommand \bibinfo  [0]{\@secondoftwo}%
\providecommand \bibfield  [0]{\@secondoftwo}%
\providecommand \translation [1]{[#1]}%
\providecommand \BibitemOpen [0]{}%
\providecommand \bibitemStop [0]{}%
\providecommand \bibitemNoStop [0]{.\EOS\space}%
\providecommand \EOS [0]{\spacefactor3000\relax}%
\providecommand \BibitemShut  [1]{\csname bibitem#1\endcsname}%
\let\auto@bib@innerbib\@empty
\bibitem [{\citenamefont {Smoluchowski}(1906)}]{Smoluchowski06}%
  \BibitemOpen
  \bibfield  {author} {\bibinfo {author} {\bibfnamefont {M.}~\bibnamefont
  {Smoluchowski}},\ }\href@noop {} {\bibfield  {journal} {\bibinfo  {journal}
  {Ann. Physik (Leipzig)}\ }\textbf {\bibinfo {volume} {21}},\ \bibinfo {pages}
  {756} (\bibinfo {year} {1906})}\BibitemShut {NoStop}%
\bibitem [{\citenamefont {Leyvraz}(2003)}]{Leyvraz03}%
  \BibitemOpen
  \bibfield  {author} {\bibinfo {author} {\bibfnamefont {F.}~\bibnamefont
  {Leyvraz}},\ }\href@noop {} {\bibfield  {journal} {\bibinfo  {journal} {Phys.
  Rep.}\ }\textbf {\bibinfo {volume} {383}},\ \bibinfo {pages} {95} (\bibinfo
  {year} {2003})}\BibitemShut {NoStop}%
\bibitem [{\citenamefont {Lushnikov}(2006)}]{Lushnikov06}%
  \BibitemOpen
  \bibfield  {author} {\bibinfo {author} {\bibfnamefont {A.~A.}\ \bibnamefont
  {Lushnikov}},\ }\href@noop {} {\bibfield  {journal} {\bibinfo  {journal}
  {Physica D}\ }\textbf {\bibinfo {volume} {222}},\ \bibinfo {pages} {37}
  (\bibinfo {year} {2006})}\BibitemShut {NoStop}%
\bibitem [{\citenamefont {Pruppacher}\ and\ \citenamefont
  {Klett}(1997)}]{Pruppacher97}%
  \BibitemOpen
  \bibfield  {author} {\bibinfo {author} {\bibfnamefont {H.~R.}\ \bibnamefont
  {Pruppacher}}\ and\ \bibinfo {author} {\bibfnamefont {J.~D.}\ \bibnamefont
  {Klett}},\ }\href@noop {} {\emph {\bibinfo {title} {Microphysics of Clouds
  and Precipitation}}}\ (\bibinfo  {publisher} {Kluwer Academic, Boston},\
  \bibinfo {year} {1997})\BibitemShut {NoStop}%
\bibitem [{\citenamefont {Mehlig}\ and\ \citenamefont
  {Wilkinson}(2004)}]{Mehlig04}%
  \BibitemOpen
  \bibfield  {author} {\bibinfo {author} {\bibfnamefont {B.}~\bibnamefont
  {Mehlig}}\ and\ \bibinfo {author} {\bibfnamefont {M.}~\bibnamefont
  {Wilkinson}},\ }\href@noop {} {\bibfield  {journal} {\bibinfo  {journal}
  {Phys. Rev. Lett.}\ }\textbf {\bibinfo {volume} {92}},\ \bibinfo {pages}
  {250602} (\bibinfo {year} {2004})}\BibitemShut {NoStop}%
\bibitem [{\citenamefont {McGraw}\ and\ \citenamefont {Liu}(2006)}]{McGraw06}%
  \BibitemOpen
  \bibfield  {author} {\bibinfo {author} {\bibfnamefont {R.}~\bibnamefont
  {McGraw}}\ and\ \bibinfo {author} {\bibfnamefont {Y.}~\bibnamefont {Liu}},\
  }\href@noop {} {\bibfield  {journal} {\bibinfo  {journal} {Geophys. Res.
  Lett.}\ }\textbf {\bibinfo {volume} {33}},\ \bibinfo {pages} {L03802}
  (\bibinfo {year} {2006})}\BibitemShut {NoStop}%
\bibitem [{\citenamefont {Spicer}\ and\ \citenamefont
  {Pratsinis}(1996)}]{Spicer96}%
  \BibitemOpen
  \bibfield  {author} {\bibinfo {author} {\bibfnamefont {P.~T.}\ \bibnamefont
  {Spicer}}\ and\ \bibinfo {author} {\bibfnamefont {S.~E.}\ \bibnamefont
  {Pratsinis}},\ }\href {\doibase 10.1002/aic.690420612} {\bibfield  {journal}
  {\bibinfo  {journal} {AIChE Journal}\ }\textbf {\bibinfo {volume} {42}},\
  \bibinfo {pages} {1612} (\bibinfo {year} {1996})}\BibitemShut {NoStop}%
\bibitem [{\citenamefont {Lushnikov}(1978)}]{Lushnikov78}%
  \BibitemOpen
  \bibfield  {author} {\bibinfo {author} {\bibfnamefont {A.~A.}\ \bibnamefont
  {Lushnikov}},\ }\href@noop {} {\bibfield  {journal} {\bibinfo  {journal} {J.
  Colloid Interface Sci.}\ }\textbf {\bibinfo {volume} {65}},\ \bibinfo {pages}
  {276} (\bibinfo {year} {1978})}\BibitemShut {NoStop}%
\bibitem [{\citenamefont {Lushnikov}(2005)}]{Lushnikov05}%
  \BibitemOpen
  \bibfield  {author} {\bibinfo {author} {\bibfnamefont {A.~A.}\ \bibnamefont
  {Lushnikov}},\ }\href@noop {} {\bibfield  {journal} {\bibinfo  {journal} {J.
  Phys. A: Math. Gen.}\ }\textbf {\bibinfo {volume} {38}},\ \bibinfo {pages}
  {L383} (\bibinfo {year} {2005})}\BibitemShut {NoStop}%
\bibitem [{\citenamefont {Krug}(2003)}]{Krug03}%
  \BibitemOpen
  \bibfield  {author} {\bibinfo {author} {\bibfnamefont {J.}~\bibnamefont
  {Krug}},\ }\href@noop {} {\bibfield  {journal} {\bibinfo  {journal} {Phys.
  Rev. E}\ }\textbf {\bibinfo {volume} {67}},\ \bibinfo {pages} {065102(R)}
  (\bibinfo {year} {2003})}\BibitemShut {NoStop}%
\bibitem [{\citenamefont {Smith}\ and\ \citenamefont
  {Matsoukas}(1998)}]{Smith98}%
  \BibitemOpen
  \bibfield  {author} {\bibinfo {author} {\bibfnamefont {M.}~\bibnamefont
  {Smith}}\ and\ \bibinfo {author} {\bibfnamefont {T.}~\bibnamefont
  {Matsoukas}},\ }\href@noop {} {\bibfield  {journal} {\bibinfo  {journal}
  {Chem. Eng. Sci.}\ }\textbf {\bibinfo {volume} {53}},\ \bibinfo {pages}
  {1777} (\bibinfo {year} {1998})}\BibitemShut {NoStop}%
\bibitem [{\citenamefont {Hendriks}\ \emph {et~al.}(1985)\citenamefont
  {Hendriks}, \citenamefont {Spouge}, \citenamefont {Eibl},\ and\ \citenamefont
  {Schreckenberg}}]{Hendriks85}%
  \BibitemOpen
  \bibfield  {author} {\bibinfo {author} {\bibfnamefont {E.~M.}\ \bibnamefont
  {Hendriks}}, \bibinfo {author} {\bibfnamefont {J.~L.}\ \bibnamefont
  {Spouge}}, \bibinfo {author} {\bibfnamefont {M.}~\bibnamefont {Eibl}}, \ and\
  \bibinfo {author} {\bibfnamefont {M.}~\bibnamefont {Schreckenberg}},\
  }\href@noop {} {\bibfield  {journal} {\bibinfo  {journal} {Z. Phys. B -
  Condensed Matter}\ }\textbf {\bibinfo {volume} {58}},\ \bibinfo {pages} {219}
  (\bibinfo {year} {1985})}\BibitemShut {NoStop}%
\bibitem [{\citenamefont {Doi}(1976{\natexlab{a}})}]{Doi76a}%
  \BibitemOpen
  \bibfield  {author} {\bibinfo {author} {\bibfnamefont {M.}~\bibnamefont
  {Doi}},\ }\href@noop {} {\bibfield  {journal} {\bibinfo  {journal} {J. Phys.
  A: Math. Gen}\ }\textbf {\bibinfo {volume} {9}},\ \bibinfo {pages} {1465}
  (\bibinfo {year} {1976}{\natexlab{a}})}\BibitemShut {NoStop}%
\bibitem [{\citenamefont {Doi}(1976{\natexlab{b}})}]{Doi76b}%
  \BibitemOpen
  \bibfield  {author} {\bibinfo {author} {\bibfnamefont {M.}~\bibnamefont
  {Doi}},\ }\href@noop {} {\bibfield  {journal} {\bibinfo  {journal} {J. Phys.
  A: Math. Gen}\ }\textbf {\bibinfo {volume} {9}},\ \bibinfo {pages} {1479}
  (\bibinfo {year} {1976}{\natexlab{b}})}\BibitemShut {NoStop}%
\bibitem [{\citenamefont {Peliti}(1985)}]{Peliti85}%
  \BibitemOpen
  \bibfield  {author} {\bibinfo {author} {\bibfnamefont {L.}~\bibnamefont
  {Peliti}},\ }\href {\doibase 10.1051/jphys:019850046090146900} {\bibfield
  {journal} {\bibinfo  {journal} {J. de Phys.}\ }\textbf {\bibinfo {volume}
  {46}},\ \bibinfo {pages} {1469} (\bibinfo {year} {1985})}\BibitemShut
  {NoStop}%
\bibitem [{\citenamefont {Patzlaff}\ \emph {et~al.}(1994)\citenamefont
  {Patzlaff}, \citenamefont {Sandow},\ and\ \citenamefont
  {Trimper}}]{Patzlaff94}%
  \BibitemOpen
  \bibfield  {author} {\bibinfo {author} {\bibfnamefont {H.}~\bibnamefont
  {Patzlaff}}, \bibinfo {author} {\bibfnamefont {S.}~\bibnamefont {Sandow}}, \
  and\ \bibinfo {author} {\bibfnamefont {S.}~\bibnamefont {Trimper}},\
  }\href@noop {} {\bibfield  {journal} {\bibinfo  {journal} {Z. f{\"u}r Phys.
  B}\ }\textbf {\bibinfo {volume} {95}},\ \bibinfo {pages} {357} (\bibinfo
  {year} {1994})}\BibitemShut {NoStop}%
\bibitem [{\citenamefont {Rey}\ and\ \citenamefont {Droz}(1997)}]{Rey97}%
  \BibitemOpen
  \bibfield  {author} {\bibinfo {author} {\bibfnamefont {P.-A.}\ \bibnamefont
  {Rey}}\ and\ \bibinfo {author} {\bibfnamefont {M.}~\bibnamefont {Droz}},\
  }\href@noop {} {\bibfield  {journal} {\bibinfo  {journal} {J. Phys. A: Math.
  Gen.}\ }\textbf {\bibinfo {volume} {30}},\ \bibinfo {pages} {1101} (\bibinfo
  {year} {1997})}\BibitemShut {NoStop}%
\bibitem [{\citenamefont {Mattis}\ and\ \citenamefont
  {Glasser}(1998)}]{Mattis98}%
  \BibitemOpen
  \bibfield  {author} {\bibinfo {author} {\bibfnamefont {D.~C.}\ \bibnamefont
  {Mattis}}\ and\ \bibinfo {author} {\bibfnamefont {M.~L.}\ \bibnamefont
  {Glasser}},\ }\href@noop {} {\bibfield  {journal} {\bibinfo  {journal} {Rev.
  Mod. Phys.}\ }\textbf {\bibinfo {volume} {70}},\ \bibinfo {pages} {979}
  (\bibinfo {year} {1998})}\BibitemShut {NoStop}%
\bibitem [{\citenamefont {Sasai}\ and\ \citenamefont
  {Wolynes}(2003)}]{Sasai03}%
  \BibitemOpen
  \bibfield  {author} {\bibinfo {author} {\bibfnamefont {M.}~\bibnamefont
  {Sasai}}\ and\ \bibinfo {author} {\bibfnamefont {P.~G.}\ \bibnamefont
  {Wolynes}},\ }\href@noop {} {\bibfield  {journal} {\bibinfo  {journal} {Proc.
  Nat. Acad. Sci. USA}\ }\textbf {\bibinfo {volume} {100}},\ \bibinfo {pages}
  {2374} (\bibinfo {year} {2003})}\BibitemShut {NoStop}%
\bibitem [{\citenamefont {T{\"a}uber}\ \emph {et~al.}(2005)\citenamefont
  {T{\"a}uber}, \citenamefont {Howard},\ and\ \citenamefont
  {{Vollmayr-Lee}}}]{Tauber05}%
  \BibitemOpen
  \bibfield  {author} {\bibinfo {author} {\bibfnamefont {U.~C.}\ \bibnamefont
  {T{\"a}uber}}, \bibinfo {author} {\bibfnamefont {M.}~\bibnamefont {Howard}},
  \ and\ \bibinfo {author} {\bibfnamefont {B.~P.}\ \bibnamefont
  {{Vollmayr-Lee}}},\ }\href@noop {} {\bibfield  {journal} {\bibinfo  {journal}
  {J. Phys. A: Math. Gen.}\ }\textbf {\bibinfo {volume} {38}},\ \bibinfo
  {pages} {R79} (\bibinfo {year} {2005})}\BibitemShut {NoStop}%
\bibitem [{\citenamefont {Ohkubo}(2007{\natexlab{a}})}]{Ohkubo07}%
  \BibitemOpen
  \bibfield  {author} {\bibinfo {author} {\bibfnamefont {J.}~\bibnamefont
  {Ohkubo}},\ }\href@noop {} {\bibfield  {journal} {\bibinfo  {journal} {J.
  Stat. Mech.: Theory Exp. {\textup P09017}}\ } (\bibinfo {year}
  {2007}{\natexlab{a}})}\BibitemShut {NoStop}%
\bibitem [{\citenamefont {Schulz}(2008)}]{Schulz08}%
  \BibitemOpen
  \bibfield  {author} {\bibinfo {author} {\bibfnamefont {M.}~\bibnamefont
  {Schulz}},\ }\href@noop {} {\bibfield  {journal} {\bibinfo  {journal} {Eur.
  Phys. J. Special Topics}\ }\textbf {\bibinfo {volume} {161}},\ \bibinfo
  {pages} {143} (\bibinfo {year} {2008})}\BibitemShut {NoStop}%
\bibitem [{\citenamefont {Ohkubo}(2007{\natexlab{b}})}]{Ohkubo11}%
  \BibitemOpen
  \bibfield  {author} {\bibinfo {author} {\bibfnamefont {J.}~\bibnamefont
  {Ohkubo}},\ }\href@noop {} {\bibfield  {journal} {\bibinfo  {journal} {Phys.
  Rev. E}\ }\textbf {\bibinfo {volume} {83}},\ \bibinfo {pages} {041915}
  (\bibinfo {year} {2007}{\natexlab{b}})}\BibitemShut {NoStop}%
\bibitem [{\citenamefont {Gardiner}\ and\ \citenamefont
  {Chaturvedi}(1977)}]{Gardiner77}%
  \BibitemOpen
  \bibfield  {author} {\bibinfo {author} {\bibfnamefont {C.~W.}\ \bibnamefont
  {Gardiner}}\ and\ \bibinfo {author} {\bibfnamefont {S.}~\bibnamefont
  {Chaturvedi}},\ }\href {\doibase 10.1007/BF01014349} {\bibfield  {journal}
  {\bibinfo  {journal} {J. Stat. Phys.}\ }\textbf {\bibinfo {volume} {17}},\
  \bibinfo {pages} {429} (\bibinfo {year} {1977})}\BibitemShut {NoStop}%
\bibitem [{\citenamefont {Gardiner}(2009)}]{Gardiner09}%
  \BibitemOpen
  \bibfield  {author} {\bibinfo {author} {\bibfnamefont {C.}~\bibnamefont
  {Gardiner}},\ }\href@noop {} {\emph {\bibinfo {title} {Stochastic Methods: A
  Handbook for the Natural and Social Sciences}}}\ (\bibinfo  {publisher}
  {Springer},\ \bibinfo {year} {2009})\BibitemShut {NoStop}%
\bibitem [{\citenamefont {Drummond}(2004)}]{Drummond04}%
  \BibitemOpen
  \bibfield  {author} {\bibinfo {author} {\bibfnamefont {P.~D.}\ \bibnamefont
  {Drummond}},\ }\href {\doibase 10.1140/epjb/e2004-00157-2} {\bibfield
  {journal} {\bibinfo  {journal} {Eur. Phys. J. B}\ }\textbf {\bibinfo {volume}
  {38}},\ \bibinfo {pages} {617} (\bibinfo {year} {2004})}\BibitemShut
  {NoStop}%
\bibitem [{\citenamefont {Drummond}\ \emph {et~al.}(2010)\citenamefont
  {Drummond}, \citenamefont {Vaughan},\ and\ \citenamefont
  {Drummond}}]{Drummond10}%
  \BibitemOpen
  \bibfield  {author} {\bibinfo {author} {\bibfnamefont {P.~D.}\ \bibnamefont
  {Drummond}}, \bibinfo {author} {\bibfnamefont {T.~G.}\ \bibnamefont
  {Vaughan}}, \ and\ \bibinfo {author} {\bibfnamefont {A.~J.}\ \bibnamefont
  {Drummond}},\ }\href {\doibase 10.1021/jp104471e} {\bibfield  {journal}
  {\bibinfo  {journal} {J. Phys. Chem. A}\ }\textbf {\bibinfo {volume} {114}},\
  \bibinfo {pages} {10481} (\bibinfo {year} {2010})}\BibitemShut {NoStop}%
\bibitem [{\citenamefont {Deloubri{\`e}re}\ \emph {et~al.}(2002)\citenamefont
  {Deloubri{\`e}re}, \citenamefont {Frachebourg}, \citenamefont {Hilhorst},\
  and\ \citenamefont {Kitahara}}]{Deloubriere02}%
  \BibitemOpen
  \bibfield  {author} {\bibinfo {author} {\bibfnamefont {O.}~\bibnamefont
  {Deloubri{\`e}re}}, \bibinfo {author} {\bibfnamefont {L.}~\bibnamefont
  {Frachebourg}}, \bibinfo {author} {\bibfnamefont {H.~J.}\ \bibnamefont
  {Hilhorst}}, \ and\ \bibinfo {author} {\bibfnamefont {K.}~\bibnamefont
  {Kitahara}},\ }\href@noop {} {\bibfield  {journal} {\bibinfo  {journal}
  {Physica A}\ }\textbf {\bibinfo {volume} {308}},\ \bibinfo {pages} {135}
  (\bibinfo {year} {2002})}\BibitemShut {NoStop}%
\bibitem [{\citenamefont {Hochberg}\ \emph {et~al.}(2006)\citenamefont
  {Hochberg}, \citenamefont {Zorzano},\ and\ \citenamefont
  {Mor\'{a}n}}]{Hochberg06}%
  \BibitemOpen
  \bibfield  {author} {\bibinfo {author} {\bibfnamefont {D.}~\bibnamefont
  {Hochberg}}, \bibinfo {author} {\bibfnamefont {M.-P.}\ \bibnamefont
  {Zorzano}}, \ and\ \bibinfo {author} {\bibfnamefont {F.}~\bibnamefont
  {Mor\'{a}n}},\ }\href@noop {} {\bibfield  {journal} {\bibinfo  {journal}
  {Chem. Phys. Lett.}\ }\textbf {\bibinfo {volume} {423}},\ \bibinfo {pages}
  {54} (\bibinfo {year} {2006})}\BibitemShut {NoStop}%
\bibitem [{\citenamefont {Gillespie}(1977)}]{Gillespie77}%
  \BibitemOpen
  \bibfield  {author} {\bibinfo {author} {\bibfnamefont {D.~T.}\ \bibnamefont
  {Gillespie}},\ }\href@noop {} {\bibfield  {journal} {\bibinfo  {journal} {J.
  Phys. Chem.}\ }\textbf {\bibinfo {volume} {81}},\ \bibinfo {pages} {2340}
  (\bibinfo {year} {1977})}\BibitemShut {NoStop}%
\bibitem [{\citenamefont {Barzykin}\ and\ \citenamefont
  {Tachiya}(2005)}]{Barzykin05}%
  \BibitemOpen
  \bibfield  {author} {\bibinfo {author} {\bibfnamefont {A.~V.}\ \bibnamefont
  {Barzykin}}\ and\ \bibinfo {author} {\bibfnamefont {M.}~\bibnamefont
  {Tachiya}},\ }\href@noop {} {\bibfield  {journal} {\bibinfo  {journal} {Phys.
  Rev. B}\ }\textbf {\bibinfo {volume} {72}},\ \bibinfo {pages} {075425}
  (\bibinfo {year} {2005})}\BibitemShut {NoStop}%
\bibitem [{\citenamefont {Biham}\ and\ \citenamefont {Furman}(2001)}]{Biham01}%
  \BibitemOpen
  \bibfield  {author} {\bibinfo {author} {\bibfnamefont {O.}~\bibnamefont
  {Biham}}\ and\ \bibinfo {author} {\bibfnamefont {I.}~\bibnamefont {Furman}},\
  }\href@noop {} {\bibfield  {journal} {\bibinfo  {journal} {Astrophys. J.}\
  }\textbf {\bibinfo {volume} {553}},\ \bibinfo {pages} {595} (\bibinfo {year}
  {2001})}\BibitemShut {NoStop}%
\bibitem [{\citenamefont {Lushnikov}\ \emph {et~al.}(2003)\citenamefont
  {Lushnikov}, \citenamefont {Bhatt},\ and\ \citenamefont
  {Ford}}]{Lushnikov03}%
  \BibitemOpen
  \bibfield  {author} {\bibinfo {author} {\bibfnamefont {A.~A.}\ \bibnamefont
  {Lushnikov}}, \bibinfo {author} {\bibfnamefont {J.~S.}\ \bibnamefont
  {Bhatt}}, \ and\ \bibinfo {author} {\bibfnamefont {I.~J.}\ \bibnamefont
  {Ford}},\ }\href@noop {} {\bibfield  {journal} {\bibinfo  {journal} {J.
  Aerosol Sci.}\ }\textbf {\bibinfo {volume} {34}},\ \bibinfo {pages} {1117}
  (\bibinfo {year} {2003})}\BibitemShut {NoStop}%
\bibitem [{\citenamefont {Bhatt}\ and\ \citenamefont {Ford}(2003)}]{Bhatt03}%
  \BibitemOpen
  \bibfield  {author} {\bibinfo {author} {\bibfnamefont {J.~S.}\ \bibnamefont
  {Bhatt}}\ and\ \bibinfo {author} {\bibfnamefont {I.~J.}\ \bibnamefont
  {Ford}},\ }\href@noop {} {\bibfield  {journal} {\bibinfo  {journal} {J. Chem.
  Phys.}\ }\textbf {\bibinfo {volume} {118}},\ \bibinfo {pages} {3166}
  (\bibinfo {year} {2003})}\BibitemShut {NoStop}%
\bibitem [{\citenamefont {Green}\ \emph {et~al.}(2001)\citenamefont {Green},
  \citenamefont {Toniazzo}, \citenamefont {Pilling}, \citenamefont {Ruffle},
  \citenamefont {Bell},\ and\ \citenamefont {Hartquist}}]{Green01}%
  \BibitemOpen
  \bibfield  {author} {\bibinfo {author} {\bibfnamefont {N.~J.~B.}\
  \bibnamefont {Green}}, \bibinfo {author} {\bibfnamefont {T.}~\bibnamefont
  {Toniazzo}}, \bibinfo {author} {\bibfnamefont {M.~J.}\ \bibnamefont
  {Pilling}}, \bibinfo {author} {\bibfnamefont {D.~P.}\ \bibnamefont {Ruffle}},
  \bibinfo {author} {\bibfnamefont {N.}~\bibnamefont {Bell}}, \ and\ \bibinfo
  {author} {\bibfnamefont {T.~W.}\ \bibnamefont {Hartquist}},\ }\href {\doibase
  10.1051/0004-6361:20010961} {\bibfield  {journal} {\bibinfo  {journal}
  {Astron. \& Astrophys.}\ }\textbf {\bibinfo {volume} {375}},\ \bibinfo
  {pages} {1111} (\bibinfo {year} {2001})}\BibitemShut {NoStop}%
\bibitem [{\citenamefont {{Losert-Valiente Kroon}}\ and\ \citenamefont
  {Ford}(2007)}]{LosertV-K07}%
  \BibitemOpen
  \bibfield  {author} {\bibinfo {author} {\bibfnamefont {C.~M.}\ \bibnamefont
  {{Losert-Valiente Kroon}}}\ and\ \bibinfo {author} {\bibfnamefont {I.~J.}\
  \bibnamefont {Ford}},\ }\href@noop {} {}\bibinfo {howpublished}
  {arXiv:0710.5540v1} (\bibinfo {year} {2007})\BibitemShut {NoStop}%
\bibitem [{\citenamefont {Yvinec}\ \emph {et~al.}(2012)\citenamefont {Yvinec},
  \citenamefont {D'Orsogna},\ and\ \citenamefont {Chou}}]{Yvinec12}%
  \BibitemOpen
  \bibfield  {author} {\bibinfo {author} {\bibfnamefont {R.}~\bibnamefont
  {Yvinec}}, \bibinfo {author} {\bibfnamefont {M.~R.}\ \bibnamefont
  {D'Orsogna}}, \ and\ \bibinfo {author} {\bibfnamefont {T.}~\bibnamefont
  {Chou}},\ }\href@noop {} {\bibfield  {journal} {\bibinfo  {journal} {J. Chem.
  Phys.}\ }\textbf {\bibinfo {volume} {137}},\ \bibinfo {pages} {244107}
  (\bibinfo {year} {2012})}\BibitemShut {NoStop}%
\bibitem [{\citenamefont {{M. Abramowitz and I. A.
  Stegun}}(1972)}]{AbramStegun}%
  \BibitemOpen
  \bibfield  {author} {\bibinfo {author} {\bibnamefont {{M. Abramowitz and I.
  A. Stegun}}},\ }\href@noop {} {\emph {\bibinfo {title} {Handbook of
  Mathematical Functions with Formulas, Graphs and Mathematical Tables}}}\
  (\bibinfo  {publisher} {Dover, New York},\ \bibinfo {year}
  {1972})\BibitemShut {NoStop}%
\bibitem [{\citenamefont {Cameron}\ and\ \citenamefont
  {Martin}(1944)}]{CameronMartin44}%
  \BibitemOpen
  \bibfield  {author} {\bibinfo {author} {\bibfnamefont {R.~H.}\ \bibnamefont
  {Cameron}}\ and\ \bibinfo {author} {\bibfnamefont {W.~T.}\ \bibnamefont
  {Martin}},\ }\href@noop {} {\bibfield  {journal} {\bibinfo  {journal} {Ann.
  Math.}\ }\textbf {\bibinfo {volume} {45}},\ \bibinfo {pages} {386} (\bibinfo
  {year} {1944})}\BibitemShut {NoStop}%
\bibitem [{\citenamefont {Cameron}\ and\ \citenamefont
  {Martin}(1945)}]{CameronMartin45}%
  \BibitemOpen
  \bibfield  {author} {\bibinfo {author} {\bibfnamefont {R.~H.}\ \bibnamefont
  {Cameron}}\ and\ \bibinfo {author} {\bibfnamefont {W.~T.}\ \bibnamefont
  {Martin}},\ }\href@noop {} {\bibfield  {journal} {\bibinfo  {journal} {Trans.
  Amer. Math. Soc.}\ }\textbf {\bibinfo {volume} {58}},\ \bibinfo {pages} {184}
  (\bibinfo {year} {1945})}\BibitemShut {NoStop}%
\bibitem [{\citenamefont {Girsanov}(1960)}]{Girsanov60}%
  \BibitemOpen
  \bibfield  {author} {\bibinfo {author} {\bibfnamefont {I.~V.}\ \bibnamefont
  {Girsanov}},\ }\href@noop {} {\bibfield  {journal} {\bibinfo  {journal}
  {Theor. Probab. Appl.}\ }\textbf {\bibinfo {volume} {5}},\ \bibinfo {pages}
  {285} (\bibinfo {year} {1960})}\BibitemShut {NoStop}%
\bibitem [{\citenamefont {{Wolfram Research, Inc.}}(2008)}]{Mathematica}%
  \BibitemOpen
  \bibfield  {author} {\bibinfo {author} {\bibnamefont {{Wolfram Research,
  Inc.}}},\ }\href@noop {} {\emph {\bibinfo {title} {Mathematica {\rm Version
  7}}}}\ (\bibinfo  {publisher} {Wolfram Research, Inc.},\ \bibinfo {year}
  {2008})\BibitemShut {NoStop}%
\bibitem [{\citenamefont {Baxter}\ and\ \citenamefont
  {Rennie}(1996)}]{BaxterRennie96}%
  \BibitemOpen
  \bibfield  {author} {\bibinfo {author} {\bibfnamefont {M.~W.}\ \bibnamefont
  {Baxter}}\ and\ \bibinfo {author} {\bibfnamefont {A.~J.~O.}\ \bibnamefont
  {Rennie}},\ }\href@noop {} {\emph {\bibinfo {title} {Financal calculus: {An}
  introduction to derivative pricing}}}\ (\bibinfo  {publisher} {Cambridge
  University Press},\ \bibinfo {year} {1996})\BibitemShut {NoStop}%
\end{thebibliography}%

\end{document}